\documentclass[acmtog, nonacm=true]{acmart}
\acmSubmissionID{257}
\usepackage{wrapfig} 
\usepackage[ruled]{algorithm2e} 

\usepackage{subcaption} 
\usepackage{siunitx} 
\setcitestyle{square,nosort}

\usepackage{booktabs} 
\usepackage{yfonts} 
\usepackage{amsmath} 
\usepackage{soul}
\usepackage{gensymb}
\usepackage{bm}
\usepackage{upgreek}
\usepackage{appendix}
\usepackage{xcolor,colortbl} 
\usepackage{array,multirow}
\usepackage{caption}
\usepackage{subcaption}
\usepackage{xr}

\usepackage{etoolbox}

\usepackage{float}
\floatstyle{plaintop}
\restylefloat{table}
\usepackage{algorithmic}

\usepackage{lipsum}
\SetAlFnt{\small}
\SetAlCapFnt{\small}
\SetAlCapNameFnt{\small}
\SetAlCapHSkip{0pt}
\usepackage{amsmath}
\usepackage{bbm}

\acmJournal{TOG}
\definecolor{comment_color}{RGB}{0 151 57}
\usepackage{mathtools}
\usepackage{algorithmic}

\definecolor{omid_color}{RGB}{0 0 0}
\definecolor{removed_material}{RGB}{250 0 50}
\definecolor{unresolved}{RGB}{250 150 0}

\definecolor{navid_color}{RGB}{50 200 100}

\begin{document}
\title{Mixed Integer Neural Inverse Design}
\author{Navid Ansari}
\affiliation{%
\institution{Max Planck Institute for Informatics}
\city{Saarbr\"ucken}
\country{Germany}
}
\email{nansari@mpi-inf.mpg.de}

\author{Hans-Peter Seidel}
\affiliation{%
\institution{Max Planck Institute for Informatics}
\city{Saarbr\"ucken}
\country{Germany}
}
\email{hpseidel@mpi-sb.mpg.de}

\author{Vahid Babaei}
\affiliation{%
\institution{Max Planck Institute for Informatics}
\city{Saarbr\"ucken}
\country{Germany}
}
\email{vbabaei@mpi-inf.mpg.de}

\renewcommand{\shortauthors}{Ansari, Seidel, Babaei}

\begin{abstract}
In computational design and fabrication, neural networks are becoming important surrogates for bulky forward simulations.
A long-standing, intertwined question is that of inverse design: how to compute a design that satisfies a desired target performance? 
Here, we show that the piecewise linear property, very common in everyday neural networks, allows for an inverse design formulation based on mixed-integer linear programming.
Our mixed-integer inverse design uncovers globally optimal or near optimal solutions in a principled manner.  
Furthermore, our method significantly facilitates emerging, but challenging, combinatorial inverse design tasks, such as material selection.
For problems where finding the optimal solution is intractable, we develop an efficient yet near-optimal hybrid approach.
Eventually, our method is able to find solutions provably robust to possible fabrication perturbations among multiple designs with similar performances. 
\end{abstract}



\begin{teaserfigure}
     \centering
     \begin{subfigure}[b]{0.3\textwidth}
         \centering
         \includegraphics[width=\textwidth]{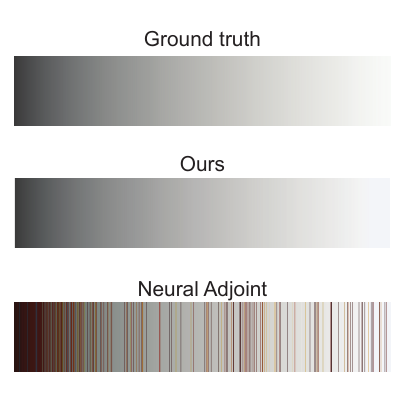}
         \caption{Spectral printing}
         \label{fig:Grayramp_teaser}
     \end{subfigure}
     \hfill
     \begin{subfigure}[b]{0.3\textwidth}
         \centering
         \includegraphics[width=\textwidth]{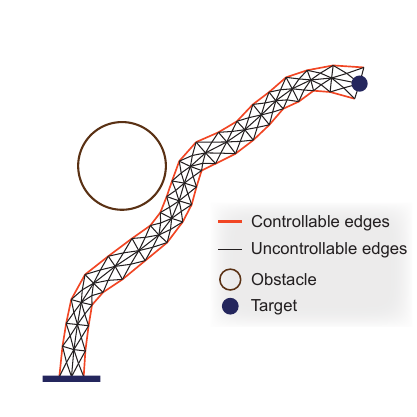}
         \caption{Soft robot inverse kinematics}
         \label{fig:softrobot_teaser}
     \end{subfigure}
     \hfill
     \begin{subfigure}[b]{0.3\textwidth}
         \centering
         \includegraphics[width=\textwidth]{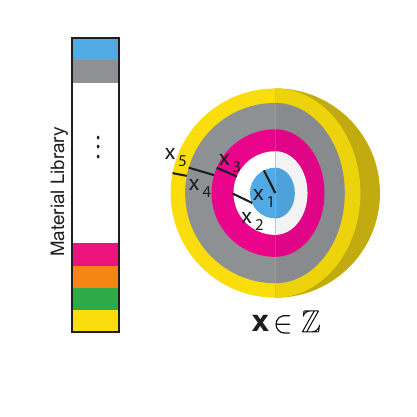}
         \caption{Nano-photonics integer inversion}
         \label{fig:Nano_sphere_teaser}
     \end{subfigure}
        \caption{We address challenging problems in neural inverse design by taking advantage of the underlying mathematical properties of neural surrogate models (NSMs). For example, in Figure~\ref{fig:Grayramp_teaser}, given a piecewise linear NSM that predicts the spectrum of a color as a function of the input ink ratios, we can find the best combination of ink ratios for reproducing a certain target spectrum. In Figure~\ref{fig:softrobot_teaser}, we find the optimal control parameters of a soft robot such that it reaches a target location while avoiding an obstacle. Figure~\ref{fig:Nano_sphere_teaser} depicts an integer-constrained inverse design problem where for a multi-shell nano-spherical scatterer we find the optimal \textit{integer} thicknesses of base materials (potentially from within a large material library) to obtain a desired scattering profile.}
        \label{fig:teaser}
\end{teaserfigure}

\maketitle
\section{Introduction} \label{intro}
Data-driven prediction of {a} design's \emph{performance} is an indispensable tool in computational {design} and fabrication. 
The amazing success of deep neural networks in computer vision and natural language processing is propelling the development of neural-network based surrogate models, or neural surrogate models (NSMs), in computational design \cite{jiang2020deep}. 
NSMs either learn and replace computationally expensive physics-based simulations \cite{kiarashinejad2020deep} or are fitted to measured data when accurate simulations are not available \cite{shi2018deep}. 
In addition to accelerating the computational design pipeline, {generality} is an implicit but important advantage of learned surrogate models: the same developed machinery can be applied to NSMs learned independently for different applications.

Forward predictions are essential for troubleshooting and analysis in computational design. 
But, oftentimes, their most important application is in \emph{inverse design}, i.e., the reverse process of mapping functional goals, or performances, into fabricable designs.
Although there has been recent progress in inverting neural networks \cite{NEURIPS2020_007ff380}, there remain many unaddressed challenges. 
Due to the non-convexity of NSMs \cite{NEURIPS2018_be53d253}, none of the current neural network inversion methods is capable of reasoning about the optimality of the obtained solutions. 
Moreover, for many naturally occurring combinatorial problems in computational design, such as {selecting} an optimal subset of materials, we still have to resort to stochastic algorithms. 

Our main insight in this work is that given a \emph{piecewise linear} neural surrogate model (PL-NSM), the inverse design problem can be formulated as a mixed-integer {linear} program (MILP).  
The piecewise linearity assumption is not particularly restrictive: most common neural networks are a composition of linear transformations, such as fully-connected or convolution layers, and piecewise linear activation functions, such as the rectified linear unit (ReLU). 
A MILP formulation of NSM-based inverse design addresses the challenges mentioned above. 
The MILP can be solved with measurable optimality as it produces the \textit{gap} between the objective's  \textit{relaxed} and \textit{feasible} solutions\footnote{\textit{Relaxed solutions} are obtained by partially dropping the integer constraints of the original problem.
As a result, in case of a minimization (maximization) the relaxed solution is an estimate guaranteed to be smaller (larger) than the optimal feasible solution.
The MILP solver tries in parallel to find better \textit{feasible solutions} and to improve the relaxed solution by progressively dropping fewer integer constraints.
When finally all constraints are considered, the relaxed and feasible solutions are equal, $gap = 0$, indicating the optimal solution is found.}. 
For small and medium sized networks, our inverse design objective typically reaches a gap of $0$, i.e., finds the globally optimal design. 
For larger networks, due to the combinatorial complexity of solving MILPs, finding global optima becomes increasingly difficult. 
Nevertheless, for these networks the solutions are still \textit{near optimal} as the {relaxed solution} can be computed via {linear} relaxation of the corresponding MILP, i.e., a linear program. 
We also show that the objective's {feasible solution} can be computed using alternative inverse methods thereby accelerating the gap closure via a hybrid of gradient-based and MILP approaches. 
Furthermore, the MILP can be straightforwardly augmented to solve challenging combinatorial inverse design problems. 
This is a significant advantage, as a large portion of inverse design problems are combinatorial by nature due to different fabrication requirements. 
Finally, when the optimization objective for many designs is similar, our method can be used to sort those solutions based on their robustness to different perturbations. 
The main contributions of this paper are: 
\begin{itemize}
    \item A novel method of inverse design via casting the inversion of piecewise linear NSMs as mixed-integer linear programming.
    \item Introducing a hybrid approach capable of providing near optimality certificate while inverting large neural networks.
    \item Proposing a framework to reliably analyze the robustness of the inverse designs. 
    \item Equipping the MILP inverse design with combinatorial constraints and applying it on a range of real-world problems.
\end{itemize}

We evaluate our proposed approaches through an extensive set of experiments in spectral printing, soft robot inverse kinematics, and photonic design. 
We will release the code to ensure the reproducibility of our results.

\section{Background and Related Work} \label{relatedwork}
\paragraph{Functional Fabrication}
One of the most important missions of computational design and fabrication is to translate functional goals, or {performances}, into fabricable designs \cite{bermano2017state}. 
In the computational fabrication literature, there are many examples trying to find a design for a prescribed performance. 
Example performances include deformation~\cite{schumacher2015microstructures}, color~\cite{sumin2019geometry}, gloss~\cite{Matusik2009}, shadow~\cite{mitra2009shadow}, relief~\cite{schuller2014appearance}, caustics~\cite{schwartzburg2014high}, etc. 
In order to solve these challenging inverse problems, often, the fabrication process is first modeled in a forward fashion where the performance is predicted from its corresponding design.
Then, to solve the original \emph{performance to design} problem, the forward process is inverted using an optimization. 
Inspired by the similarity among these problems, \citet{chen2013spec2fab} propose a framework, called \emph{spec}2\emph{fab}, that abstracts the functional fabrication process in a general manner. 
In this work, we focus on functional fabrication problems whose forward modeling can be expressed via a piecewise linear neural network.

\paragraph{Neural Networks and Computational Design}
Neural networks can map designs to performances by approximating complex physics simulations \cite{kiarashinejad2020deep}.  
Moreover, they can operate as purely data-driven simulations when accurate physics-based models are unavailable or difficult to develop \cite{shi2018deep}. 
In addition to accelerating the computations, neural surrogate models are highly transferable across different applications due to their underlying similarities. 
%
Perhaps \emph{photonic design} is the front-runner field in using neural surrogate models for computational design \cite{jiang2020deep}. 
Also, in computational fabrication we are seeing a surge in the use of NSMs, such as in computational design of cold-bent glass façades \cite{gavriil2020computational}, appearance-preserving tactile design \cite{tymms2020appearance}, or fine art reproduction \cite{shi2018deep}. 
%
A particularly relevant related work is the recent ink selection method \cite{ansari2020mixed}. 
In order to benefit from efficient MILP solvers, for the ink selection problem, this work develops a linear, but approximate, forward model that predicts the spectra of different ink combinations. 
(In general, developing such linear spaces requires deep domain knowledge and specialized measurements.) 
In a second stage, for spectral reproduction of a given painting, an accurate data-driven forward model based on NSMs is deployed.  
Here we show that \textit{both} ink selection and spectral reproduction can be performed using a single neural-network forward model without requiring additional, specialized models. 
%

\paragraph{Neural Network Inversion}
Recently, there has been a surge in inverse models for neural networks. 
The first solution coming to mind is to train neural networks in the reverse direction using the training data. 
This naive approach fails because of the one-to-many nature of the problem: the same performance could lead to different designs causing problems during optimization (e.g., via inconsistent gradients). 
In order to bypass this challenge, \emph{tandem} networks \cite{liu2018training,shi2018deep} map performances into designs using a first neural network but, in order to compute a consistent loss, use a pre-trained forward network to map the resulting design into its corresponding performance. 
Conditional variational auto-encoders \cite{kingma2013auto} have also been used for the inverse design task \cite{kiarashinejad2020deep}. 
These networks condition the design on the target performance and yield a distribution of solutions from which multiple samples could be drawn. 
An invertible neural network \cite{ardizzone2018analyzing}, based on \textit{real NVP} \cite{DBLP:conf/iclr/DinhSB17}, is another inversion method. 
In this method, a specialized architecture based on normalizing flows is trained in both forward and inverse directions leading to a bijective mapping between design and performance spaces. 
\citet{NEURIPS2020_007ff380} benchmarked these inverse methods and found out that a gradient-based method, via backpropagation with respect to the design variables, results in significantly more accurate solutions. 
Dubbed as \textit{neural adjoint} (NA), this method uses a {boundary loss} to punish infeasible designs. 
It also runs the optimization starting from multiple {random} initial guesses in search for the best objective value. 
Our method, by nature, is similar to the NA as both are \textit{optimizations} and not inverse architectures as in~\cite{liu2018training,shi2018deep,ardizzone2018analyzing}. 
In Section~\ref{sec:experiments}, we evaluate our method against the NA extensively.

\paragraph{Mixed-Integer Programming and Neural Networks}
The mathematical optimization problems in which all or some variables are integers are known as mixed-integer programming\footnote{We recommend the following short and gentle introduction to MIP and its solvers to the less familiar reader: https://www.gurobi.com/resource/mip-basics/} (MIP) \cite{floudas1995nonlinear}. 
A technique for solving MIPs with nonlinear, nonconvex functions, dating back to \citet{markowitz1957solution}, is to {estimate} those functions with piecewise linear functions \cite{belotti2013mixed}. 
The resulting approximation, usually via auxiliary binary variables, is a mixed-integer {linear} programming with more scalable and efficient solvers. 
The connection between piecewise linearity of some class of neural networks and MILP solvers has only recently been identified \cite{cheng2017maximum}. 
MILP formulation has since been exploited for formal verification of networks against adversarial attacks \cite{fischetti2018deep}. 
It is important to note that unlike the classic use of piecewise linear functions for \emph{approximating} non-linear functions, MILP representation is simply a \emph{reformulation} of the already piecewise linear networks without any approximation. 
We borrow the MILP formulation of piecewise-linear networks, initially appeared in the formal verification literature \cite{fischetti2018deep,NEURIPS2018_be53d253,tjeng2018evaluating}, and develop a novel neural inverse design framework. 
To the best of our knowledge, we are the first to introduce the MILP-based \textit{neural inverse design} and extend it to related tasks, such as combinatorial inverse design problems.

\section{Neural Inverse Design via Mixed-Integer Linear Programming} \label{sec:method}
In this section, we take a forward model expressed as a trained, piecewise linear neural network and invert it using mixed-integer linear programming. 
In addition to solving typical inverse design problems, we show how extra integer constraints can readily be added to our pipeline allowing for solving challenging combinatorial inverse design problems. 
Additionally, this formulation can be easily adapted for evaluating the robustness of different designs.
Finally, our proposed method can be combined with other inversion methods in order to find more accurate near-optimal designs efficiently. 
%
%

\subsection{Mixed-Integer Formulation} \label{sec:formulation}
A \emph{feedforward} neural network $F_{\theta}$ is built by a number of function compositions \cite{montufar2014number}

\begin{equation} \label{eq:nndefinition}
\begin{split}
     \mathbf{x}^{L} = F_{\theta}(\mathbf{x^{0}})=f^{L} \circ g^{L-1} \circ f^{L-1} \circ \ldots \circ g^{1} \circ f^{1} (\mathbf{x}^{0}) \\ 
    \end{split}
\end{equation}

\noindent 
and maps the input $\mathbf{x}^{0} \in \mathbb{R}^m$ to the output $\mathbf{x}^{L} \in \mathbb{R}^n$ (note that the last layer does not undergo an activation). 
Here, $f^{l}$ is a linear pre-activation function 

\begin{equation} 
\begin{split}
     f^{l} (\mathbf{x}^{l-1}) = \mathbf{W}^{l} \mathbf{x}^{l-1}+\mathbf{b}^{l} \\ 
     \quad \forall \;  l \in \left \{ 1,2, \cdots, L \right \}
         \end{split}
\end{equation}
\noindent 
whose weights $\mathbf{W}^{l}$ and biases $\mathbf{b}^{l}$ at all layers (1 to $L$) make up the network's parameters $\theta$ which are computed during the training. 
In our notation superscripts and subscripts (to appear later) indicate the layers and nodes, respectively. 
The function $g^{l}$ is a nonlinear activation function. 
Throughout all inverse problems in this work we assume the widely used rectified linear unit or ReLU as the activation function. 
Using other piecewise linear activation functions, such as leaky ReLU or max pooling layers is {also} straightforward.
The ReLU function is defined as 
\begin{equation}
    \mathbf{x}^{l} = g^{l} (f^{l} (\mathbf{x}^{l-1})) = \operatorname{max}  \{\bm{0},\mathbf{W}^{l} \mathbf{x}^{l-1}+\mathbf{b}^{l} \}. 
    \label{eq:reludefinition}
\end{equation}
\noindent
We adopt a vector-matrix notation for compactness and readability. 
That is, the $\mathrm{max}$ operator in Equation~\ref{eq:reludefinition} takes a vector input and outputs the component-by-component maxima. 

In a general neural inverse problem we search for an input vector $\mathbf{x^{0}}$ that minimizes a distance, using $\mathcal{L}_{1}$ norm\footnote{The $\mathcal{L}_{1}$ norm is more amenable to linearization.}, between the network prediction and a target performance $\mathbf{t}$
\begin{equation}
\underset{\mathbf{x}^{0}}{\operatorname{argmin}} \left \| F_{\theta}(\mathbf{x^{0}}) - \mathbf{t}  \right \|_{1}.
\label{eq:generalminimization}
\end{equation}
\noindent
This optimization is very challenging as $F_{\theta}$ is a highly non-linear, non-convex function \cite{NEURIPS2018_be53d253}. 
Nevertheless, we can exploit the piecewise linear structure of neural networks and model their optimization using mixed-integer linear programming. 
That is, the optimization only involves linear terms and constraints.
In doing so, we eliminate the network's non-linearities at the cost of introducing new {binary} and continuous variables.

We adapt the MILP-based reformulation of ReLU networks \cite{tjeng2018evaluating} (previously used for formal verification) for solving our central inversion problem, summarized in Equation~\ref{eq:generalminimization}.  
Given a pretrained network $F_{\theta}$ with a given set of weights $\mathbf{W}^{l}$ and biases $\mathbf{b}^{l}$, we encode the inverse problem shown in Equation~\ref{eq:generalminimization} as a MILP with linear and binary constraints\footnote{Note the curled inequalities and $\odot$ symbol indicate our continuing use of component-by-component convention.}

\begin{subequations}
\begin{equation}
\underset{\mathbf{z}^{1}, \ \cdots \ , \mathbf{z}^{L-1},\quad \mathbf{x}^{0}, \ \cdots \ , \mathbf{x}^{L}}{\operatorname{argmin}} \left \| \mathbf{x}^{L} - \mathbf{t}  \right \|_{1}
\label{eq:milptargetminimization}
\end{equation}
\begin{equation}
\mathbf{x}^{l} \preceq  \mathbf{W}^{l} \mathbf{x}^{l-1} + \mathbf{b}^{l}  \; - \;  \mathbf{l}^{l} (\bm{1} - \mathbf{z}^{l}) 
\label{eq:lower_bound}
\end{equation}
\begin{equation}
\mathbf{x}^{l} \succeq   \mathbf{W}^{l} \mathbf{x}^{l-1}+\mathbf{b}^{l}  
\label{eq:LargerThanInput}
\end{equation}
\begin{equation}
\mathbf{x}^{l} \preceq  \mathbf{u}^{l}  \odot  \mathbf{z}^{l}
 \label{eq:upperbound}
 \end{equation}
\begin{equation}
\mathbf{x}^{l} \succeq  \bm{0} 
\label{eq:LargerThanZero}
\end{equation}
\begin{equation}
\mathbf{z}^{l} \in \{0,1\}^{K^{l}} .
\label{eq:binary_Z}
\end{equation}
\label{eq:relumilp}
\end{subequations}

\noindent
For the {nodes} at layer $l$ we introduce a set of continuous ($\mathbf{x}^{l}$) and binary ($\mathbf{z}^{l}$) variables. 
Vectors $\mathbf{l}^{l}$ and $\mathbf{u}^{l}$ are the lower and upper bounds to the nodes' pre-activation values $\mathbf{W}^{l} \mathbf{x}^{l-1}+ \mathbf{b}^{l}$ and are precomputed (see Section~\ref{sec:tightening}). 

While optimizing Equation~\ref{eq:relumilp}, the solver \textit{branches} on these binary variables and, at the worst case, checks all possible network's configurations.
It is simple to verify that the constraints replace the role of $\operatorname{max} \{0,.\}$ operation: when $\mathbf{z}^l_k=1$ (corresponding to neuron $k$ in layer $l$), the constraints \ref{eq:lower_bound} and \ref{eq:LargerThanInput} are binding and thus $\mathbf{x}^{l} =  \mathbf{W}^{l} \mathbf{x}^{l-1} + \mathbf{b}^{l}$ for the neuron $k$. 
Otherwise, when $\mathbf{z}^l_k=0$, the constraints in equations \ref{eq:upperbound} and \ref{eq:LargerThanZero} are binding and thus $\mathbf{x}^l_k=0$.
Note that while we are mostly interested in the optimized value of $\mathbf{x}^{0}$, we should optimize all introduced binary and continuous variables to enforce the constraints in Equation~\ref{eq:relumilp} and thus have a correct representation of the neural network. 
%

\subsection{Combinatorial Inverse Design} \label{sec:combinDesign}
The MILP representation of the inverse design problem can take additional integer constraints in a seamless manner. 
These integer constraints appear in many computational design problems. 
For example, in a \emph{selection} problem, we are interested in a limited number $D$ of all input design features $\mathbf{x}^0$, which results in an optimal target performance $\mathbf{t}$. 
We can cast the selection problem as Equations~\ref{eq:milptargetminimization} to \ref{eq:binary_Z} \textbf{in addition} to

\begin{subequations}
\begin{equation}
\sum_{i=1}^{K^0} \mathbf{q}_{i} \leq D  
\label{eq:sum_to_d}
\end{equation}
\begin{equation}
\mathbf{q} \in \{0,1\}^{K^{0}} 
\label{eq:binary_q}
\end{equation}
\begin{equation}
    0  \leq  \mathbf{x}_{i}^0  \leq  \mathbf{q}_{i},  \; \; \; \;     \forall \;  i \in \left \{ 1,2, \cdots, K^0 \right \}
\label{eq:q_and_x}
\end{equation}

\label{eq:selectionconstraints}
\end{subequations}

\noindent
where the vector of inputs to the neural network $\mathbf{x}^0$ is of size $K^0$ and normalized between $0$ and $1$, and $\mathbf{q}$ is our introduced \emph{selection variable}, a {binary} vector of same size (different from previously defined binary variables $\mathbf{z}^l$). 
The inequality constraints \ref{eq:sum_to_d} and \ref{eq:q_and_x} ensure that at most $D$ entries of $\mathbf{x}^0$ take non-zero values and thus used for estimating $\mathbf{t}$. 
Indices of these entries match the indices of non-zero elements in $\mathbf{q}$ and {point to the selected elements}. 
Other combinatorial inverse design problems can be formulated similarly by adding proper constraints and integer variables. 
In Sections~\ref{sec:milpnnselection} and \ref{sec:resultintegerinversedesign} we show how this formulation is applied to real-world inverse problems.

\subsection{Bound Precomputation} \label{sec:tightening}

\begin{wrapfigure}[14]{r}{0.14\textwidth}
  \vspace{-4.5mm}
  \centering
  \hspace{-10.5mm}
  \includegraphics[width=0.14\textwidth]{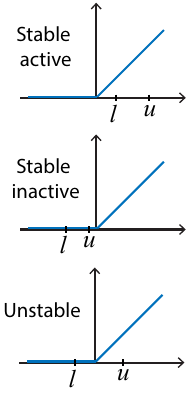}\\
  \vspace{-1.5mm}
  \label{fig:relutypes}
\end{wrapfigure}

We precompute as tight as possible lower $\mathbf{l}^{l}$ and upper $\mathbf{u}^{l}$ bounds to the pre-activation values $\mathbf{W}^{l} \mathbf{x}^{l-1}+ \mathbf{b}^{l}$. 
There are two main, interrelated advantages in bound tightening. 
First, it improves the solve time of the problem by strengthening its formulation \cite{vielma2015mixed}. 
Second, tighter bounds can lead to more \textit{stable} ReLUs. 
Stable ReLUs are those that operate on nodes whose bounds lie completely within either positive or negative domain (see the inset).
When the bounds lie within the positive domain (stably active), the value of such a node is always a linear combination of preceding nodes and there is no need to introduce {new} optimization variables for it. 
When the bounds lie within the negative domain (stably inactive), the value of such a node is always zero and therefore the corresponding variables are dropped. 
Otherwise, we have \textit{unstable} ReLUs for which we {must} define binary and continuous variables.

The procedure for bound precomputation is similar to our original inverse problem where we calculate the minimum (maximum) of each node of the neural network using the same mixed-integer formulation (Equation~\ref{eq:relumilp}). 
Except that instead of minimizing (maximizing) the distance to the target $\mathbf{t}$, we minimize (maximize) the value of each individual node $k$ in layer $l$: 

\begin{equation}
\underset{\mathbf{z}^{1}, \ \cdots \ , \mathbf{z}^{l-1}, \mathbf{z}^{l}_{k}, \quad \mathbf{x}^{0}, \ \cdots \ , \mathbf{x}^{l-1}, \mathbf{x}^{l}_{k}}{\operatorname{argmin}}  \mathbf{x}^{l}_{k}
\label{eq:lbOptimization}
\end{equation}

\begin{equation}
(\underset{\mathbf{z}^{1}, \ \cdots \ , \mathbf{z}^{l-1}, \mathbf{z}^{l}_{k}, \quad \mathbf{x}^{0}, \ \cdots \ , \mathbf{x}^{l-1}, \mathbf{x}^{l}_{k}}{\operatorname{argmax}}  \mathbf{x}^{l}_{k}).
\label{eq:ubOptimization}
\end{equation}

\noindent
This optimization is still subjected to constraints ~\ref{eq:lower_bound} to \ref{eq:binary_Z} for the considered \textit{node} and all preceding layers.

\begin{wrapfigure}[16]{r}{0.20\textwidth}
  \vspace{-4.5mm}
  \centering
  \hspace{-10.5mm}
  \includegraphics[width=0.20\textwidth]{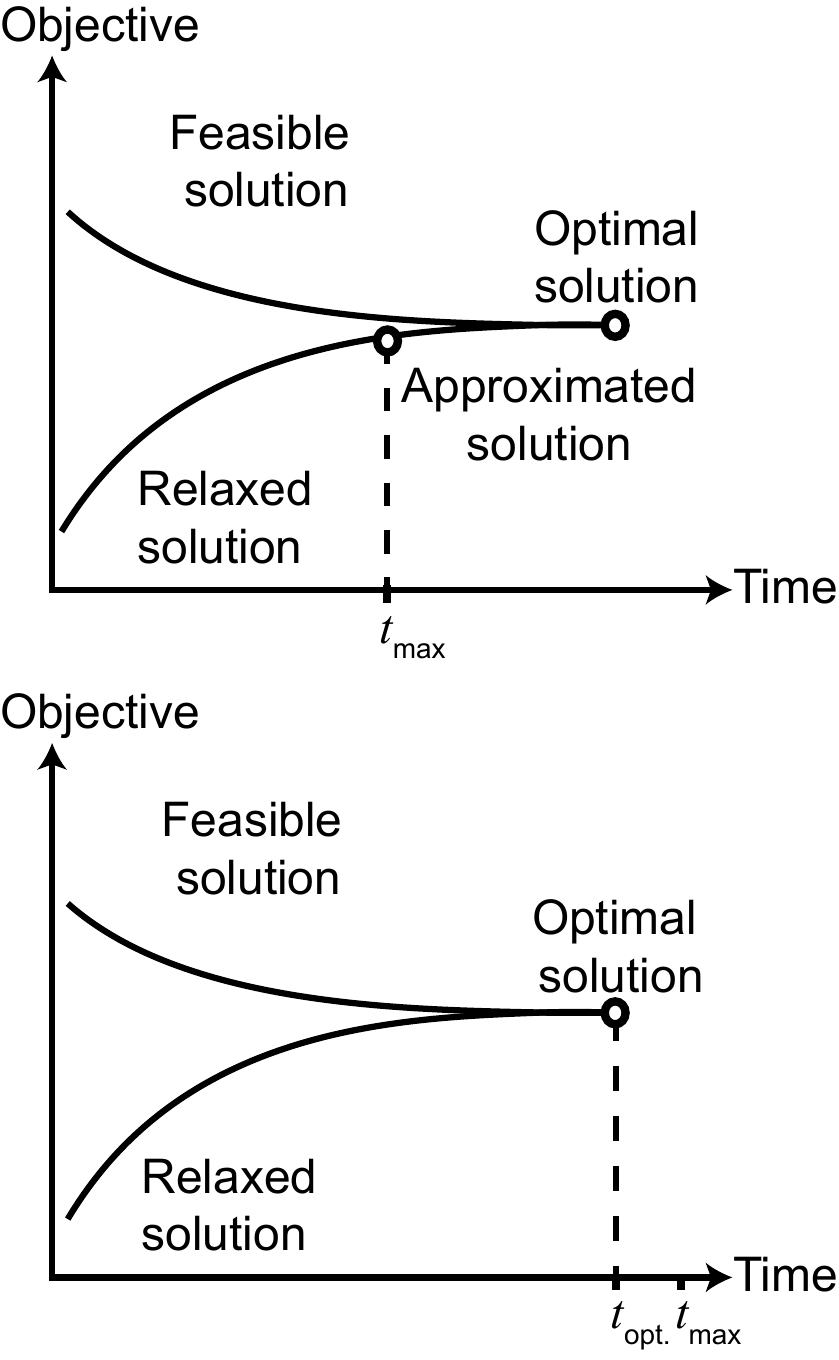}\\
  \vspace{-1.5mm}
  \label{fig:boundscheme}
\end{wrapfigure}

Our bound tightening algorithm is based on \cite{fischetti2018deep} but extended to include the design constraints.
Since designs ($\mathbf{x^{0}}$ in our notation) usually come with their own constraints, we observe that it is highly beneficial to enforce those constraints when precomputing the nodes' bounds as they lead to tighter bounds.
The bound precomputation can be very expensive for larger networks as it should be performed on each node separately. 
The computation is especially heavy within last layers of the network. 
In practice, we set a time limit ({$t$\textsubscript{max}}) for the solver during this computation. 
{If we stop the optimization prematurely, the relaxed solution of the optimization is the node's bound.
As depicted in the inset figures, relaxed solutions are conservative estimation of the optimal solutions and guarantee that in case of a minimization (maximization) there are no smaller (larger) solutions than this estimation.
The feasible solution, however, is the solution that is found for the original MILP problem thus far and there could be smaller (larger) values if the minimization (maximization) continues.
It is important to note that using feasible solution as upper and lower bounds in an early-stopped optimization, results in overestimating the lower bound and underestimating the upper bound values.
This will lead to calculating incorrectly tighter bounds and overestimating the number of stable ReLUs, which results in sub-optimal solutions.
{Algorithm~\ref{alg:precomputation} in Appendix~\ref{sec:alg_appendix} presents the extended bound precomputation step by step.}
}

\subsection{Combination of Gradient-Based Optimizations and MILP} \label{sec:NaMilp}
For non-combinatorial inverse design problems, gradient-based optimizations are an attractive choice given that the network's gradient information can be efficiently computed via automatic differentiation. 
The neural adjoint (NA) method \cite{NEURIPS2020_007ff380}, for example, relies on a gradient descent approach based on backpropagation algorithm \cite{hecht1992theory} for inverting neural networks. 
The process is very similar to network training except instead of network's parameters its input is optimized. 
Despite its good scalability with network's size, this (and any other) neural inverse method is unable to reason about the optimality of the obtained solutions.

Using our method, once the neural inverse design is formulated via MILP, we immediately obtain a {relaxed solution} to the objective. 
Typical MILP solvers search recursively for both {feasible and relaxed solutions} of the objective and try to close their gap as quickly as possible \cite{klotz2013practical}. 
Therefore, any feasible solution is a near optimal solution because we know how far it is from the (conservative) {relaxed solution} at any moment.  
Given the obtained solutions via NA are all feasible to the {MILP} objective, we can solve for the {relaxed solution} of the objective via MILP and its {feasible solution} via NA. 
{In practice we run NA and the MILP on the same network simultaneously and track the optimality gap.
We can then stop the process by monitoring the gap. 
Note that, for larger neural networks (in our experience approximately larger than 4 layers each 150 neuron wide) it is not tractable to insist on a zero gap.}
As we show in Section~\ref{sec:namilpexperiments}, NA solutions (in case of minimization) reduce the objective's {feasible solutions} significantly more quickly resulting in a tighter objective's bound in a less amount of time.

\subsection{Design Robustness} \label{sec:robustnesstheory}
Inverse designs are typically one-to-many problems where for a given performance there are multiple acceptable designs. %
It is therefore interesting to study other attributes during inverse design. 
An important attribute is the robustness of designs to possible perturbations during fabrication \cite{sigmund2009manufacturing}. 
We define the robustness of a computed, candidate design $\hat{\mathbf{x}}^{0}$ as the maximum deviation of its performance from a desired target performance $\mathbf{t}$ when the candidate design is perturbed by a small positive number $\epsilon$ at each dimension. 
In other words, we look for the worst performance of a design when it is allowed to roam inside a {hypercube} around it.  
The mixed integer formulation allows us to find the \textit{provably} worst performance. 
We write this problem as 
\begin{equation}
\begin{split}
\underset{\mathbf{z}^{1}, \ \cdots \ , \mathbf{z}^{L-1},\quad \mathbf{x}^{0}, \ \cdots \ , \mathbf{x}^{L}}{\operatorname{argmax}} \left \| \mathbf{x}^{L} - \mathbf{t}  \right \|_{1} \label{eq:robustness} \\ 
\hat{\mathbf{x}}^{0}_{i} - \epsilon \leqslant \mathbf{x}^{0}_{i} \leqslant \hat{\mathbf{x}}^{0}_{i} + \epsilon, \quad \forall \;  i \in \left \{ 1,2, \cdots, K^{0} \right \}. 
\end{split}
\end{equation}

\noindent
{Once again this optimization is subjected to constraints ~\ref{eq:lower_bound} to \ref{eq:binary_Z}. 
}
Note that the candidate design $\hat{\mathbf{x}}^{0}$ need not necessarily come from MILP-based inversion. 
In our case, in Section~\ref{sec:robustnessexperiments}, we use the neural adjoint (NA) method for computing candidate designs. 
In general, robustness computation is more efficient than typical MILP-based inversion as the design is usually perturbed within a small neighborhood.
This, on top of bound precomputation, leads to further reduction of unstable ReLUs. 

\section{Evaluation} \label{sec:experiments}
In this section, we demonstrate the potential of our proposed method. 
For our analyses and experiments, we focus on several real-life applications in three different areas of computational design and control: neural spectral printing \cite{shi2018deep,ansari2020mixed}, inverse kinematic of soft robots \cite{xue2020amortized, sun2021amortized}, and photonic design \cite{peurifoy2018nanophotonic,nadell2019deep}. 
We solve all MILPs using Gurobi, a state-of-the-art solver \cite{gurobi}. 
{In order to better relate the experiments to the theory (developed in 
Section \ref{sec:method}), in Table \ref{tab:Variable_describtion}, we summarize the setup of each experiment in connection with Equations \ref{eq:relumilp} and \ref{eq:selectionconstraints}.} 
All bound precomputation and MILPs, unless otherwise mentioned, are solved on a CPU cluster with 256 cores. 
This does not mean that employing all cores is always desirable when solving a MILP.
In practice, we find that using more than 30 cores does not help with the speed up.
On the other hand, the nodes' bound precomputation is trivially parallelizable for the nodes belonging to the same layer.  
The time limit $t$\textsubscript{max} for bound precomputation is set to 150 seconds. 
In {all} experiments, the reported error is the objective value of our optimization (based on the $\mathcal{L}_{1}$ norm) for a set of unseen, target performances. 

\subsection{Neural Spectral Separation} \label{sec:resultinversedesign}

\begin{figure}
         \centering
         \includegraphics[width=0.5\textwidth]{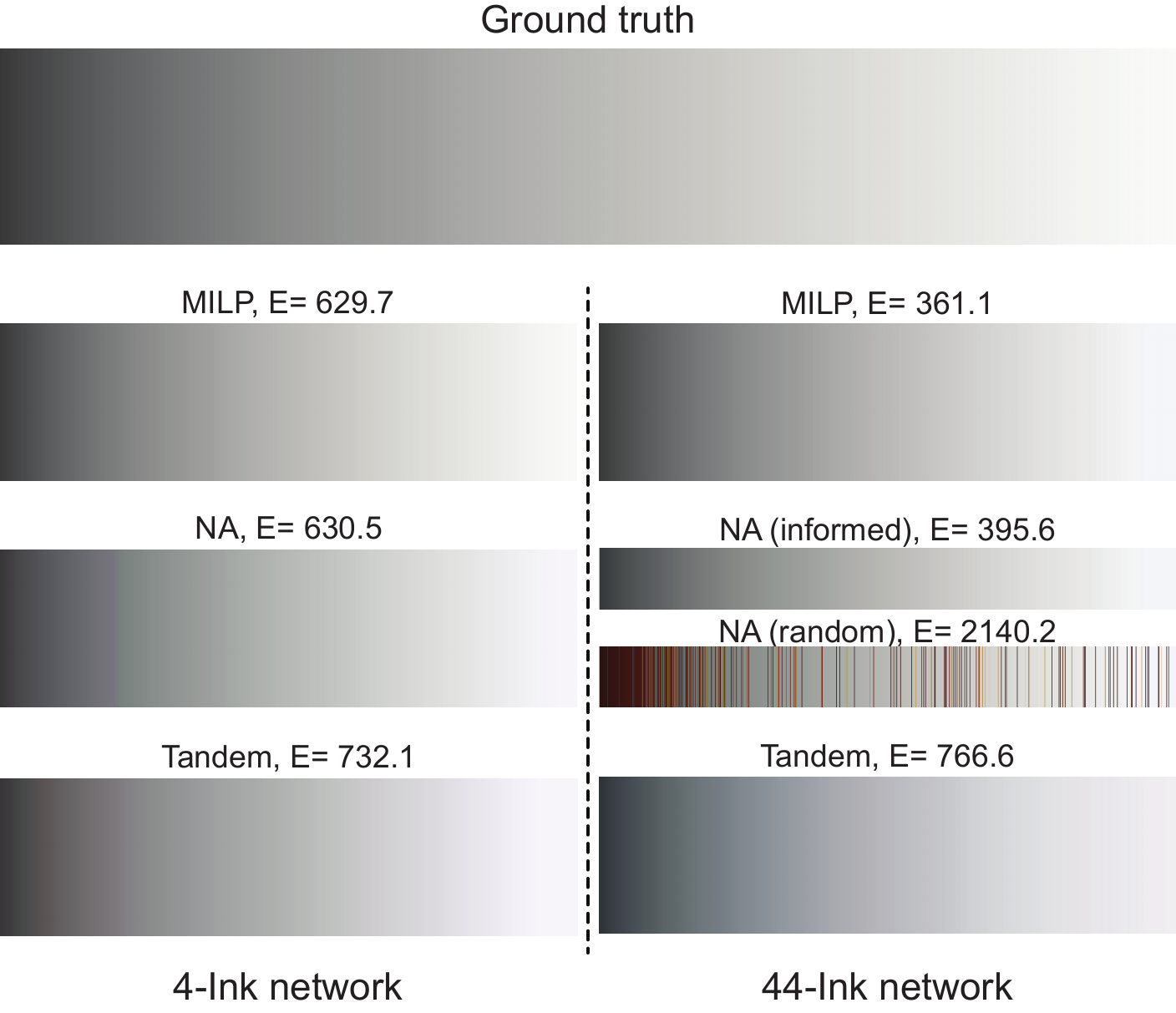}

        \caption{Different neural inverse methods for spectral separation of a perfect gray ramp. The error (E) is the sum of objective for all 901 gray spectra. We split the NA's solution in the middle to show both random, and domain-knowledge informed initializations. }
        \label{fig:grayramp}
\end{figure}

We begin by studying a neural inverse problem in spectral printing.
Spectral printing ensures that printed items are visually close to the originals, independent of the color of the light source under which they are observed. 
{In this experiment we used two different neural networks as the input to our method.
The 4-ink network is a trained PL-NSM with 4 hidden layers each having 100 neurons and ReLU activation functions \cite{ansari2020mixed}.
The 44-ink network is a trained PL-NSM with 2 hidden layers each having 50 neurons and ReLU activation functions.} 
{The 4-ink network}, a surrogate for a forward spectral prediction model, has a 4D design space made of CMYK (cyan, magenta, yellow and black) ink ratios and outputs a 31D spectrum. 
Here the inverse design problem, known as \textit{spectral separation}, concerns finding the ink ratios for a target spectrum.
A particularly challenging target for spectral separation is the perfect gray ramp introduced by \citet{ansari2020mixed}. 
This gradient is formed by $901$ dark to light ideal gray spectra which have equal reflectively across all visible wavelengths (Figure~\ref{fig:grayramp}).

We compare our MILP-based inversion with both the method of tandem, previously used for the exactly same problem \cite{shi2018deep,ansari2020mixed}, and the neural adjoint (NA) \cite{NEURIPS2020_007ff380} as it has been shown to significantly outperform other neural inverse methods in literature. 
Figure~\ref{fig:grayramp} visualizes the accuracy of different inverse methods for spectral separation. 
Our method performs $901$ separate optimizations (i.e., Equation~\ref{eq:relumilp}) in order to find the corresponding ink ratios. 
For NA, we run the $901$ optimizations, each $50$ times with different random initialization. 
We allow {up to} {2000} iterations of Adam  ~\cite{kingma2014adam} and terminate the optimization if the solution does not improve within a threshold in 10 consecutive iterations. 
In the tandem method we query the learned inverse method {once} using {a batch of $901$} gray spectra as input. 

Looking at the spectral separation accuracies in Figure~\ref{fig:grayramp}, using the 4-ink PL-NSM, both our method and NA perform very well surpassing the tandem method significantly. 
In fact, with a gap of $0$ between the {feasible and relaxed solution} of the objective, our method finds the \emph{global} optima for all 901 instances. 
This indicates that the method of NA has also performed remarkably well as its error is only slightly worse than our method. 
For a better comparison, we re-run an identical set of experiments using a new PL-NSM with 44 inks as input, i.e., a 44D design space. 
For this network, our method again finds the global optima for all spectral targets. 
This time, however, NA struggles to find acceptable solutions for many targets and produces an erroneous reproduction.  
This indicates that NA's performance drops for higher dimensional design spaces likely due to random initialization.

In a second experiment on NA, instead of random initialization, we initialize the optimization with an \textit{informed} guess. 
That is, in a crude estimation, we assume that the average spectra of all input inks, i.e., when each ink's contribution is $1/44$, is a 50\% gray spectrum. 
Therefore, for reproducing the pure black, i.e., the darkest gray (100\%), each ink is initialized with $2/44$ ratio, and so on. 
With this informed initialization drawn from domain knowledge, the accuracy of NA increases substantially but still trails MILP's performance. 
{This experiment reveals a significant advantage of our method where unlike NA, due to global optimality, increasing the design space dimensionality does not affect the accuracy.} 
Being insensitive to the initialization is another major advantage as using domain knowledge for informed initial guesses is not always feasible.

Using the MILP approach, the optimization of a single gray spectrum takes on average around $256$ and $840$ seconds for the 4- and 44-ink PL-NSM, respectively.  
We spend also around $331$ and $3.5$ seconds on a one-time precomputation of upper and lower bounds of the network nodes.
The {NA} method, on a Titan X GPU, takes on average $62$  and $99$ seconds for the 4- and 44-ink PL-NSM, respectively. 
The fastest method is tandem, {spending less than 1 second for all spectra}, as querying neural network is extremely efficient, though at the cost of significant accuracy loss. 
%


\definecolor{lg}{gray}{0.92}
\definecolor{lgg}{gray}{0.975}
\newcolumntype{g}{>{\columncolor{lg}}c}
	\begin{table*}
		\centering
		\resizebox{\linewidth}{!}{%
			\begin{tabular}{cgggggggggggggg} 
				\toprule
				& \multicolumn{5}{c}{Variables of Equations~\ref{eq:relumilp} and/or \ref{eq:selectionconstraints}}\\
				\cmidrule(lr){3-5}
				\rowcolor{white}
				Experiment name & Experiment mode & $\mathbf{t}$ (Target) & $\mathbf{x}^0$ (Input)& $\mathbf{x}^L$ (Output) & Design constraints& Variable of interest\\
				\toprule


\emph{Neural Spectral}& 4-ink network & Perfect gray  & Ink ratios & Gray spectrum&  $0\leq\mathbf{x}^0 \leq 1$& $\mathbf{x}^0$ \\

\emph{Separation}& & spectrum ($\mathbf{t} \in \mathbb{R}^{31}$) &  ($\mathbf{x}^0 \in \mathbb{R}^{4}$)&  ($\mathbf{x}^L \in \mathbb{R}^{31}$) & &  \\

\rowcolor{white}
 &44-ink network& Perfect gray & Ink ratios & Gray spectrum & $0\leq\mathbf{x}^0 \leq 1$ & $\mathbf{x}^0$\\
 \rowcolor{white}
  &  & spectrum ($\mathbf{t} \in \mathbb{R}^{31}$)&  ($\mathbf{x}^0 \in \mathbb{R}^{44}$) &  ($\mathbf{x}^L \in \mathbb{R}^{31}$) &  &\\

\cmidrule(lr){1-7}
 \rowcolor{white}
 \emph{Soft Robot}& - & Target location & Stretch and & Location of all  & Equations & $\mathbf{x^{0}}$  \\
 
\rowcolor{white}
\emph{Inverse Kinematic}& & ($\mathbf{t} \in \mathbb{R}^{2}$)& contraction ($\mathbf{x}^0 \in \mathbb{R}^{40}$)& vertices ($\mathbf{x}^{L} \in \mathbb{R}^{206}$)& \ref{eq:SoftRobot_Design_const}, \ref{eq:SoftRobot_smmothness_const}, and \ref{eq:SoftRobot_obstacle_const} & \\
\cmidrule(lr){1-7}

\emph{Material Selection}& Selection & Matrix of 6 & Matrix of 6 area & Matrix of 6 & Equation \ref{eq:selectionconstraints}, $D = 2$,&  $\mathbf{q}$ \\

&  & spectra ($\mathbf{t} \in \mathbb{R}^{6 \times 31}$) &  coverage ($\mathbf{x}^0 \in \mathbb{R}^{6 \times 44}$)& spectra ($\mathbf{x}^{L} \in \mathbb{R}^{6 \times 31}$)& $\mathbf{q} \in \left \{0, 1  \right \}^{44}$ &\\

\rowcolor{white}
 & Inversion & Painting's color & Ink ratios& Color spectrum& $0\leq\mathbf{x}^0 \leq \mathbf{q}$& $\mathbf{x}^0$ \\ 
 \rowcolor{white}
& & spectrum ($\mathbf{t} \in \mathbb{R}^{31}$) & ($\mathbf{x}^0 \in \mathbb{R}^{44}$) &($\mathbf{x}^{L} \in \mathbb{R}^{31}$)& &\\ 

\cmidrule(lr){1-7}

\emph{Nano-Photonics}& Inversion (rounded) & Scattering cross & Spherical shell &  Scattering cross&  $0\leq 10 \mathbf{x}^0 \leq 70$& $Round(\mathbf{x}^0)$\\

&  & section ($\mathbf{t} \in \mathbb{R}^{200}$)& thickness ($\mathbf{x}^0 \in \mathbb{R}^{4}$)&  section ($\mathbf{x}^{L} \in \mathbb{R}^{200}$)& &\\ 
\rowcolor{white}
& Integer-constrained & Scattering cross & Spherical shell &  Scattering cross&  $0\leq 10 \mathbf{x}^0 \leq 70$& $\mathbf{x}^0$\\
\rowcolor{white}
& inversion & section ($\mathbf{t} \in \mathbb{R}^{200}$)& thickness ($\mathbf{x}^0 \in \mathbb{Z}^{4}$)&  section ($\mathbf{x}^{L} \in \mathbb{R}^{200}$)& &\\ 
 
\cmidrule(lr){1-7}

\emph{Contoning}& Inversion (rounded) & Color spectrum & Ink layer thickness & Color spectrum&  $0 \leq \mathbf{x^{0}} \leq 30$,& $Round(\mathbf{x}^0)$\\

& & ($\mathbf{t} \in \mathbb{R}^{31}$)& ($\mathbf{x}^0 \in \mathbb{R}^{11}$)& ($\mathbf{x}^{L} \in \mathbb{R}^{31}$)&  $\sum_{1}^{11} x_{i}^{0} = 30$&\\
\rowcolor{white}
& Integer-constrained & Color spectrum & Ink layer thickness & Color spectrum&  $0 \leq \mathbf{x^{0}} \leq 30$,& $\mathbf{x}^0$\\
\rowcolor{white}
& inversion & ($\mathbf{t} \in \mathbb{R}^{31}$)& ($\mathbf{x}^0 \in \mathbb{Z}^{11}$)& ($\mathbf{x}^{L} \in \mathbb{R}^{31}$)&  $\sum_{1}^{11} x_{i}^{0} = 30$&\\
 
 \cmidrule(lr){1-7}
\rowcolor{white}
 \emph{MILP-NA}& - & Perfect gray & Ink ratios & Gray spectrum&   $0\leq\mathbf{x}^0 \leq 1$& $\mathbf{x}^0$,\\
 \rowcolor{white}
  \emph{combination}&  & spectrum ($\mathbf{t} \in \mathbb{R}^{31}$) &  ($\mathbf{x}^0 \in \mathbb{R}^{8}$)&  ($\mathbf{x}^L \in \mathbb{R}^{31}$) & & Optimality Gap \\
  
  \cmidrule(lr){1-7}
 \rowcolor{white}
 \emph{Robustness Analysis}& - & Metasurface & $4\times$cylinder height & Metasurface spectrum & $0 \leq \mathbf{x^{0}} \leq 1$,& ${\operatorname{argmax}} \left \| \mathbf{x}^{L} - \mathbf{t}  \right \|_{1}$  \\
 
  \rowcolor{white}
&  & spectrum ($\mathbf{t} \in \mathbb{R}^{300}$)& and radius ($\mathbf{x}^0 \in \mathbb{R}^{8}$)& ($\mathbf{x}^{L} \in \mathbb{R}^{300}$)& $\hat{\mathbf{x}}^{0}_{i} - 10^{-3} \leqslant \mathbf{x}^{0}_{i} \leqslant \hat{\mathbf{x}}^{0}_{i} + 10^{-3}$& \\

				\bottomrule
			\end{tabular}
		}
		\caption{The setup of each experiment in connection with Equations \ref{eq:relumilp} and \ref{eq:selectionconstraints}.}
		\label{tab:Variable_describtion}
	\end{table*}

\subsection{Soft Robot Inverse Kinematics} \label{sec:soft_robot}
{Soft robotics is dedicated to studying robots made with flexible materials, with applications in minimally invasive surgeries~\cite{majidi2014soft}, prosthetics~\cite{polygerinos2015soft} and many more. 
To control the movement of soft robots accurately, an efficient inverse kinematics approach is required. 
In this section, using our MILP approach, we solve the \textit{neural} inverse kinematics problem for soft robots, introduced by~\citet{sun2021amortized}. }
The problem involves controlling the soft robot such that it reaches a certain target location.
Following~\cite{sun2021amortized}, we consider a snake-like robot made of 103 vertices connected with flexible edges and a fixed bottom (Figure~\ref{fig:softrobot_teaser}). 
Among these edges, 40 side edges (colored in Figure~\ref{fig:softrobot_teaser}) are controllable whose stretches and contractions form our design space (40D).
This problem is typically formulated as a PDE-constrained optimization where the design parameters are the boundary conditions, and the solution of the optimization is the position of all the vertices (final shape)~\cite{xue2020amortized}.
As PDE-constrained optimizations are computationally costly, resorting to NSMs is an attractive alternative. 

We train a PL-NSM with two hidden layers, each having 128 neurons and ReLU activation functions. 
To create the training data we solve a PDE-constrained optimization for 50,000 randomly generated stretches and contractions of controllable edges~\cite{xue2020amortized}.
The model's input is the stretches and contractions of 40 controllable edges of the soft robot.
The output is the $(x, y)$ coordinates of all 103 vertices, hence, our performance space has 206 dimensions.
For the inversion, we would like to optimize the input such that the center of the soft robot's tip reaches the target location $\mathbf{t}$, i.e.,
\begin{equation}
\begin{split}
\underset{\mathbf{z}^{1}, \ \cdots \ , \mathbf{z}^{L-1},\quad \mathbf{x}^{0}, \ \cdots \ , \mathbf{x}^{L}}{\operatorname{argmin}} \left \| \mathbf{x}^{L}_{i} - \mathbf{t}  \right \|_{1} , \\
i \in \left[123, \ 124  \right],
\label{eq:SoftRobotObj}
\end{split}
\end{equation}

\noindent
where, in our setup, indices 123 and 124 represent the $(x, y)$ location of the robot's tip.
Note that Equation~\ref{eq:SoftRobotObj} is the neural inverse problem objective function (Equation~\ref{eq:milptargetminimization}) with a slight modification and is still under the constraints~\ref{eq:lower_bound} to \ref{eq:binary_Z}.

To enforce valid designs, we limit the contractions and stretches ($\mathbf{\zeta}$) in the input, 
\begin{equation}
\mathbf{\zeta_{min}} \preceq  \mathbf{x}^{0} \preceq \mathbf{\zeta_{max}}.
\label{eq:SoftRobot_Design_const}
\end{equation}
The minimum ($\mathbf{\zeta_{min}}$) and maximum ($\mathbf{\zeta_{max}}$) contractions and stretches are among soft robot properties.
Following~\citet{sun2021amortized} we set $\mathbf{\zeta_{min}}$ and $\mathbf{\zeta_{max}}$ to $-0.2$ and $0.2$, respectively.
Moreover, to avoid non-physical changes, we use a similar term to the one proposed by \citet{sun2021amortized} which controls the smoothness of the contractions and stretches in consecutive edges: 
\begin{equation}
\begin{split}
\left |  (\mathbf{x}^{0}_{i+1}-\mathbf{x}^{0}_{i}) - (\mathbf{x}^{0}_{i}-\mathbf{x}^{0}_{i-1})  \right |\leq l_{d} , \\ 
i\in (1, n), \ i\neq \frac{n}{2}, \ i\neq \frac{n}{2}+1.
\end{split}
\label{eq:SoftRobot_smmothness_const}
\end{equation}

\noindent
Here $l_{d}$ is the limit for the deformation and should be determined based on the mechanical properties of the soft robot (flexibility, stress tolerance, etc.).
In this experiment, we assume a material that can tolerate a predefined amount of deformation $l_{d}=0.2$.
Indices $i\in \left [1, n/2  \right ]$ and $i\in \left [n/2 +1, n \right ]$ correspond to the edges on the robot's left-hand side and right-hand side, respectively. 
Moreover, {$i=n/2$ and $i=n/2 + 1$ are the end-nodes from two different sides and excluded from Equation \ref{eq:SoftRobot_smmothness_const} as we apply the smoothness terms on each side separately}.

Finally, to have a more challenging setup, we seek robot paths that avoid a circular obstacle with radius $r$.
Unlike~\citet{sun2021amortized}, we use an $\mathcal{L}_{1}$ norm for avoiding the obstacle. 
\begin{figure*}[t]
         \centering
         \includegraphics[width=0.8\textwidth]{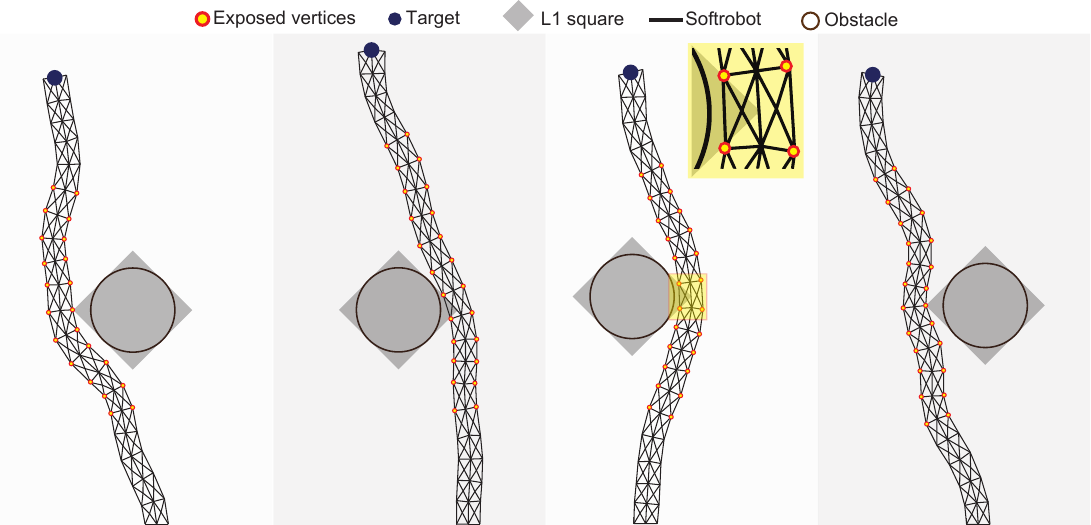}
        \caption{Soft robot inverse kinematics with MILP. Using the $\mathcal{L}_{1}$ norm square prevents the collision between robot's \textit{vertices} and the square while significantly reducing the chance of the collision between the \textit{edges} and the (primary) circular obstacle. The highlighted vertices (in blue) are those who are considered in the no-collision term.}
        \label{fig:Soft_robot_kinematic_milp}
\end{figure*}
Using $\mathcal{L}_{1}$ norm, the vertices avoid the auxiliary, \textit{square} obstacle and thus the chance of intersection between the \textit{edges} and the primary, \textit{circular} obstacle will be reduced significantly.
To avoid the obstacle, we define the following constraint
\begin{equation}
\begin{split}
\sqrt{2} r \leq  \left \|  \mathbf{x}_{i}^{L} -  \mathbf{o} \right \|_{1}
\end{split}
\label{eq:SoftRobot_obstacle_const}
\end{equation}

\noindent
where $\mathbf{o}$ is the location of the center of the obstacle, $r$ is the obstacle radius, and $\sqrt{2} r$ is {half} the diagonal of the square encompassing the obstacle.
In this experiment we set $r = 0.9$.
Note that only a subset of vertices are exposed to the obstacle {(highlighted via blue dots in Figure \ref{fig:Soft_robot_kinematic_milp})}, and if they do not collide with the obstacle, the rest of the vertices remain intact.
In Equation~\ref{eq:SoftRobot_obstacle_const}, index $i$ represents the set of such exposed vertices.

We solve this inverse problem using our MILP approach for 1000 randomly sampled target locations. 
We spend 307 seconds on a one-time network's bound pre-computation.
In average, the solve time for each sample is 9.2 seconds.
All solutions avoid the obstacle and reach $\mathcal{L}_{1}$ error of $0$.
Figure~\ref{fig:Soft_robot_kinematic_milp} shows 4 randomly chosen solution samples of the MILP optimization.
In the first three instances, it seems that the soft robot collides with the $\mathcal{L}_{1}$ square but a closer look reveals that the vertices, as expected (and guaranteed), are intact.
Note that, following the previous work, here we define the constraints based only on vertices, but designing linear constraints for edges to avoid the obstacle is also conceivable.
As mentioned earlier, using the $\mathcal{L}_{1}$ error reduces the chance of the intersection between the edges and the circular obstacle.

We also solve the neural inverse kinematics problem using the NA method.
We use the optimization objectives introduced in~\cite{sun2021amortized} with slight changes such that the losses are comparable with the MILP method ( Appendix~\ref{sec:NA_softrobot_appendix}). 
{We observe that in the NA-based inversion, all samples avoid the obstacle too, but the average $\mathcal{L}_{1}$ error is 0.0308 with the average calculation time of 84.5 seconds}.
Unlike our method, in NA different constraints such as obstacle avoidance and physical deformations are incorporated as \textit{soft} constraints in the energy term with weight hyperparameters to be tuned. 
This reveals other important advantages of the MILP formulation. 
First, different constraints can be added to the MILP simply and are guaranteed to be satisfied. 
Second, no hyperparameter tuning is required. 
The latter is especially beneficial in real-world problems where there might be many constraints that need to be incorporated in the objective function.
By introducing more constraints, it becomes significantly harder to tune the corresponding weights.
\subsection{Material Selection} \label{sec:milpnnselection}


Although digital fabrication technologies, such as multi-material 3D printers, have a limited number of channels, there is a vast array of materials that can fill those channels. 
Consequently, the question of which subset of materials is optimal for a given task (also known as {material selection}) is becoming a recurrent question \cite{ansari2020mixed,piovarvci2020towards,nindel2021gradient}.

{Here we reproduce the results of the duotone reproduction experiment from \citet{ansari2020mixed} via our approach.
The effect of the ink selection is highly visible in duotone experiment and the smallest mistake will stand out prominently. 
}
Similar to us, \citet{ansari2020mixed} employ a MILP formulation for the ink selection. 
However, they need to develop a custom \textit{linear} forward model that requires deep domain knowledge and specialized measurements.   
Interestingly, for the actual spectral separation, they train NSMs for the selected inks.  
Here we show that both spectral separation and the ink selection can be performed via purely data-driven NSMs.

\begin{figure}
         \centering
     \begin{subfigure}[b]{0.23\textwidth}
         \centering
         \includegraphics[width=\textwidth]{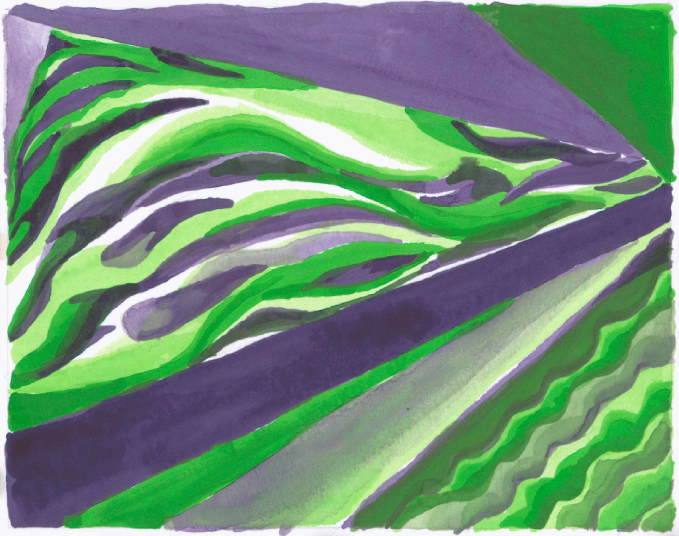}
        \caption{Original painting}
         \label{fig:Original_painting}
     \end{subfigure}
\hfill     
          \begin{subfigure}[b]{0.23\textwidth}
         \centering
         \includegraphics[width=\textwidth]{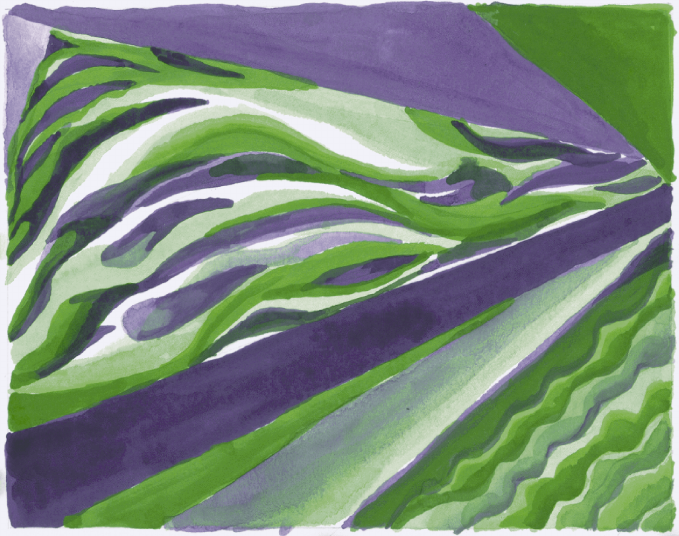}
        \caption{Reproduction}
         \label{fig:Reproduced_painting}
     \end{subfigure}
\hfill

        \caption{Given a PL-NSM that predicts the spectrum of a set of 44 inks, our method finds the optimal pair of inks that results in the best spectral reproduction of the input image.} %
\label{fig:MILP_painting_selection}
\end{figure}

In our duotone reproduction setup, following \citet{ansari2020mixed}, given a spectral image we look for the best pair of inks leading to optimal spectral reproduction from within an ink library of 44 inks. 
The input is the spectral image of a limited palette watercolor painting from \citet{ansari2020mixed} shown in Figure~\ref{fig:Original_painting}. 
We adopt the PL-NSM (Section~\ref{sec:resultinversedesign}) that predicts the spectrum of a set of 44 library inks. 
As printed data for training a 44-ink network is not provided, we simulate such data using the Neugebauer model \cite{yule1967principles}, an analytical spectral prediction model. 
The Neugebauer primaries are computed using the multiplication of library ink transmittances.
Since we are looking for a pair of inks that performs well on the whole image, following \citet{ansari2020mixed}, we sample {6} spectra from the input image. 
We use the method explained in Section~\ref{sec:combinDesign}, relying on both Equations~\ref{eq:relumilp} and \ref{eq:selectionconstraints}. 
As we use 6 sampled spectra from the input image, 
we optimize for the {6} targets $\mathbf{t}$ {simultaneously} via using a sum in Equation~\ref{eq:milptargetminimization}.
{In fact one can see this problem as solving 6 copies of the network simultaneously for 6 different target spectra all of which must satisfy Equation \ref{eq:selectionconstraints}.}
{This means that the variables are almost 6x more than solving for a single target (see Table~\ref{tab:Variable_describtion}).}

Our MILP-based ink selection finds the ground truth inks with a gap of $0$, i.e., provably the optimal pair of inks for reproduction of the given input (Figure~\ref{fig:Original_painting}). 
Having obtained the two optimal inks, in order to reproduce the input image, we could calibrate a new NSM using the reliable, printed data of those two inks.  
More interesting is to use the same 44-ink network, this time in a spectral separation configuration, in order to simulate how the pair of optimal inks reproduce the painting (Figure~\ref{fig:Reproduced_painting}).  
As we see in Figure~\ref{fig:MILP_painting_selection}, the reproduction is of high quality. 
The quality can be still improved if we calibrate the network with printed rather than simulated data. 
But the remarkable fact of this experiment remains the globally optimal ink selection on a neural network. 
The time for solving the ink selection problem is {4998s}. 
The time for the spectral separation is, on average, 1.8 seconds {for each spectra}, and manageable as there are only 5483 unique colors in the scene. 
Finally, the one-time precomputation of nodes' bounds took {3.5 seconds.}

\subsubsection{Comparison with Genetic Algorithm} 
When selecting 2 out of 44 inks, it might be tempting to try a brute-force approach where the objective for each possible pair of inks is computed and then the pair with the best objective is selected. 
There are however two major caveats. 
First, although the number of two-ink combinations in a set of 44 inks is reasonable, selecting a larger number of inks via a brute force approach is infeasible. 
For example, selecting 10 inks amounts to around $2 \times 10^9$ combinations, which means we need to perform this number of optimizations to find the objective for each combination. 
Second, in the absence of a MILP approach, the objective values may not be optimal.

\begin{figure}
     \centering
     \begin{subfigure}[b]{0.23\textwidth}
         \centering
         \includegraphics[width=\textwidth]{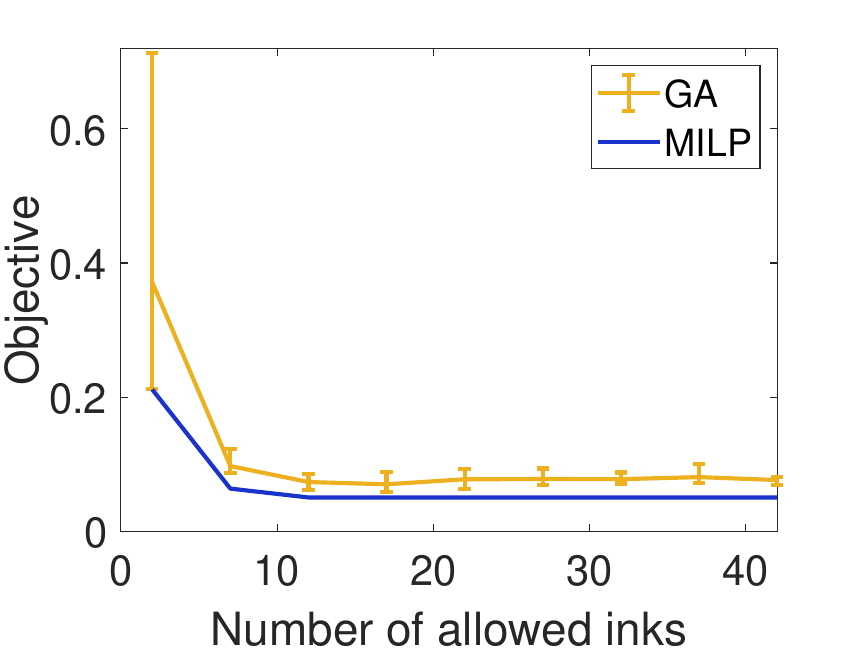}
        \caption{}
         \label{fig:GA_milp_loss}
     \end{subfigure}
\hfill     
          \begin{subfigure}[b]{0.23\textwidth}
         \centering
         \includegraphics[width=\textwidth]{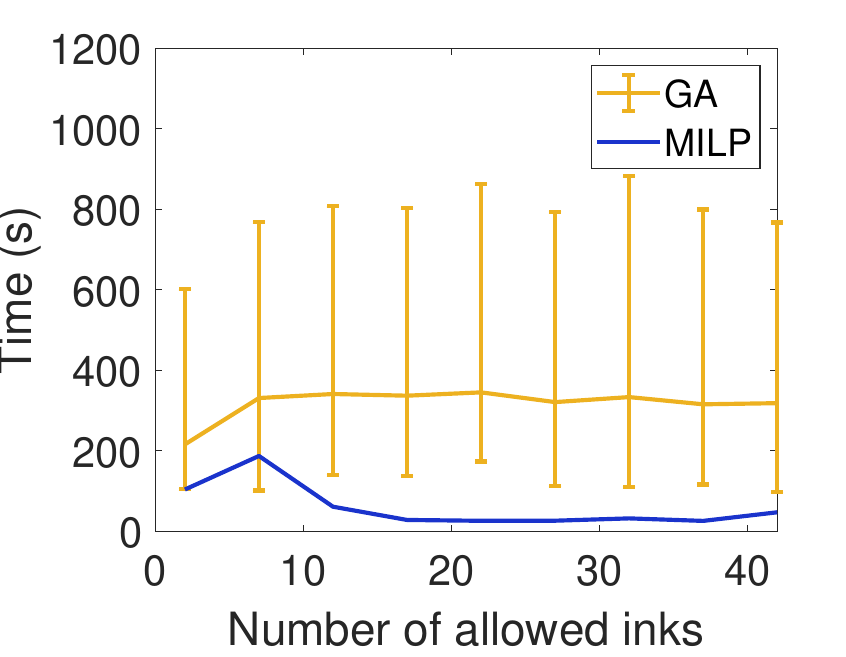}
        \caption{}
         \label{fig:GA_milp_time}
     \end{subfigure}
\hfill

        \caption{Comparing our MILP and a GA approach for an ink selection problem. We repeat GA 10 times and report the average, maximum and minimum values.} %
        \label{fig:ga_milp_performance}
\end{figure}

More appropriate solutions to such combinatorial problems are based on stochastic methods, such as genetic algorithm (GA) or simulated annealing. 
In this section we compare our proposed method to GA for a selection problem. 
Genetic algorithm searches the combinatorial space stochastically via their well-known heuristics and, in general, prefer combinations with best objectives. 
For computing the GA's objective, we use the interior point method. 
In Figure~\ref{fig:ga_milp_performance} we perform ink selection for a single target spectrum each time allowing for a different number of inks.
At each step we repeat the experiment 10 times to capture the variance in GA solutions, and show the average, maximum and minimum values. 
As shown in Figure~\ref{fig:ga_milp_performance}, the MILP approach always yields the optimal solutions, outperforming GA in both time and accuracy.
{Note that MILP is considerably faster even though minimum GA computation time on the plot seems to be smaller at around 8 inks. This is because MILP computation time should be compared to the GA's multiple run times until GA converges to a desired solution (due to stochasticity).}
In this experiment these two methods were evaluated on the same hardware (Intel Xeon CPU E5).

\subsection{Integer-Constrained Inverse Design} \label{sec:resultintegerinversedesign}
Apart from material selection, a significant portion of inverse design problems are combinatorial by nature due to the fabrication constraints. 
For example, \emph{metamaterials} are usually made of juxtaposition of a set of materials (including the void) in 2D or 3D arrays \cite{bertoldi2017flexible}. 
Current approaches assign continuous material properties (such as permittivity) to the elements of these arrays and quantize these values before fabrication. 
Unfortunately, the quantization step can significantly undo the optimized performance \cite{zhu2020inverse}.
When the forward model is expressed via a {PL-NSM}, our combinatorial inverse design formulation can seamlessly take such integer constraints into account. 
Here, we demonstrate two examples of {integer-constrained} inverse designs.  
%
\subsubsection{Nano-Photonics}

\begin{figure}
     \centering
          \begin{subfigure}[b]{0.4\textwidth}
         \centering
         \includegraphics[width=\textwidth]{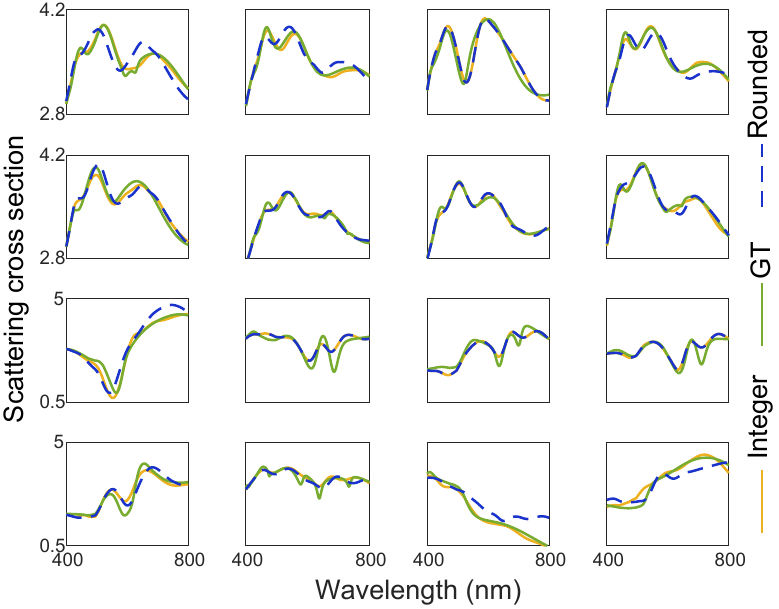}
         \caption{} 
         \label{fig:sphere_inversion}
     \end{subfigure}
\hfill    

     \centering
          \begin{subfigure}[b]{0.4\textwidth}
         \centering
         \includegraphics[width=\textwidth]{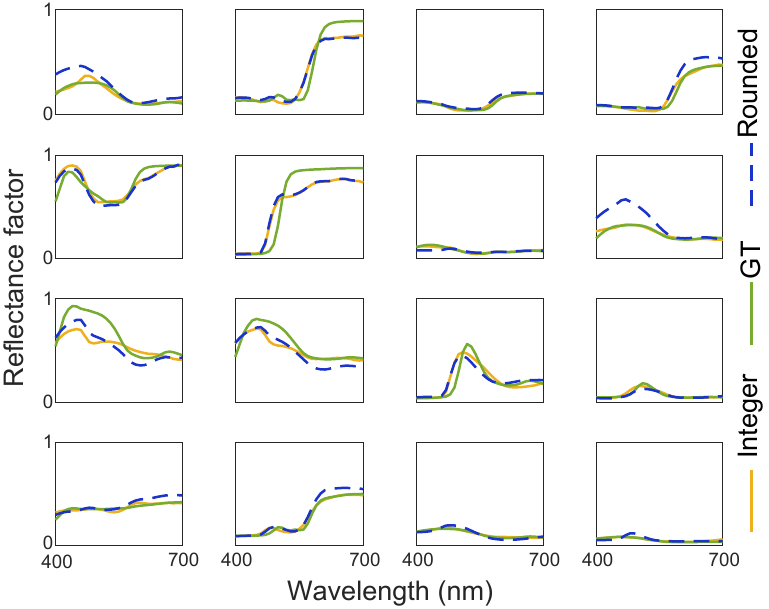}
         \caption{} 
         \label{fig:contoning_inversion}
     \end{subfigure}
\hfill  

        \caption{(a) Optimizing for integer shell thickness of a photonic nano-sphere, and (b) integer 3D printed layer thickness in contoning. In both experiments, one can ignore the integer constraints and solve for continuous solutions to be rounded to nearest integers. Our method allows for directly optimizing the desired integer values with significant accuracy gain over the rounded solution.} %
        \label{fig:integer_inversion}
\end{figure}

In this experiment (Figure~\ref{fig:Nano_sphere_teaser}), we consider the light scattering from a multilayer dielectric spherical nano-particle \cite{peurifoy2018nanophotonic}. 
We obtain a different spectral scattering cross section by changing the thickness of the material of each shell.  
Similar to spectral printing, here we also look for optimal ratios (thicknesses) of the materials which result in a desired spectrum. 
In order to imitate possible fabrication constraints, we slightly twist the experiment by limiting the materials to take a predefined set of integer thicknesses (from {0} to {70} nm at 10 nm intervals). 
We train a {PL-NSM} with {3} layers of {100, 50, and 100} neurons following \citet{peurifoy2018nanophotonic} which maps the combination of {4} materials into the resulting spectrum. 

We test our methods on 16 target scattering cross sections shown in Figure~\ref{fig:sphere_inversion} and, for all targets, reach the globally optimal solution with {average} {objective} of $19.47$. 
For comparison, we also perform the same inverse design {(via MILP)} on the targets but without enforcing integer thicknesses.
After rounding the obtained optimal but continuous thicknesses to the nearest allowed integers, the error increases significantly ({$30.30$}). 
For this particular problem, our MILP-based integer-constrained inversion takes on average {$44$} seconds. 
We also spend {$4.8$} seconds on the one-time bound precomputation of the network nodes.

\subsubsection{Contoning}
In color reproduction for 3D printing via \textit{contoning} \cite{babaei2017color}, inks with different thicknesses can be superposed. 
Contoning avoids potential artifacts of the alternative halftoning techniques that rely on spatial multiplexing of materials on the surface. 
While contoning works well when ink concentrations are low, with highly concentrated inks the quantization artifacts start to appear as the thickness can only be controlled via tuning a discrete number of layers. 
In an experiment similar to the previous one, we show how our method obtains the best discrete arrangement of different ink layers for reproducing a given spectrum.  

Following the setup of \citet{shi2018deep}, we want to reproduce a target spectrum by superposing 30 layers of 11 different inks.
We train a PL-NSM with 3 layers of 50 neurons which maps the layer layouts to the spectrum. 
%
{Our printer can print 30 layers of 11 different inks (see Table \ref{tab:Variable_describtion} for design constraints). We have performed this experiment with two different settings, in our first attempt (similar to \cite{shi2018deep}) we set the  $\mathbf{x^{0}} \in \mathbb{R}^{11}$.
Since the smallest amount of ink that our fabrication device can deposit is a single layer, we have to round the elements of $\mathbf{x^{0}}$ to the nearest integer neighbors. This rounding step introduces error to our designs.
In the second setup, by defining $\mathbf{x^{0}} \in \mathbb{Z}^{11}$ in our MILP formulation we directly solved the integer inverse problem and found the optimal integer design.
}
In Figure~\ref{fig:contoning_inversion}, we plot 16 target and reproduced spectra.
We also show the resulting spectra obtained by rounding the {optimal} continuous layer thicknesses to the closest integers. 
The {average} error in this experiment is $1.14$ and $1.72$ for optimal integer and rounded solutions, respectively. 
Our MILP-based integer-constrained spectral separation takes on average 40 seconds. 
The bound precomputation {for the considered network takes 8 seconds.}

\subsection{Combination of MILP and NA}\label{sec:namilpexperiments}
\begin{figure}
     \centering
         \centering
         \includegraphics[width=0.5\textwidth]{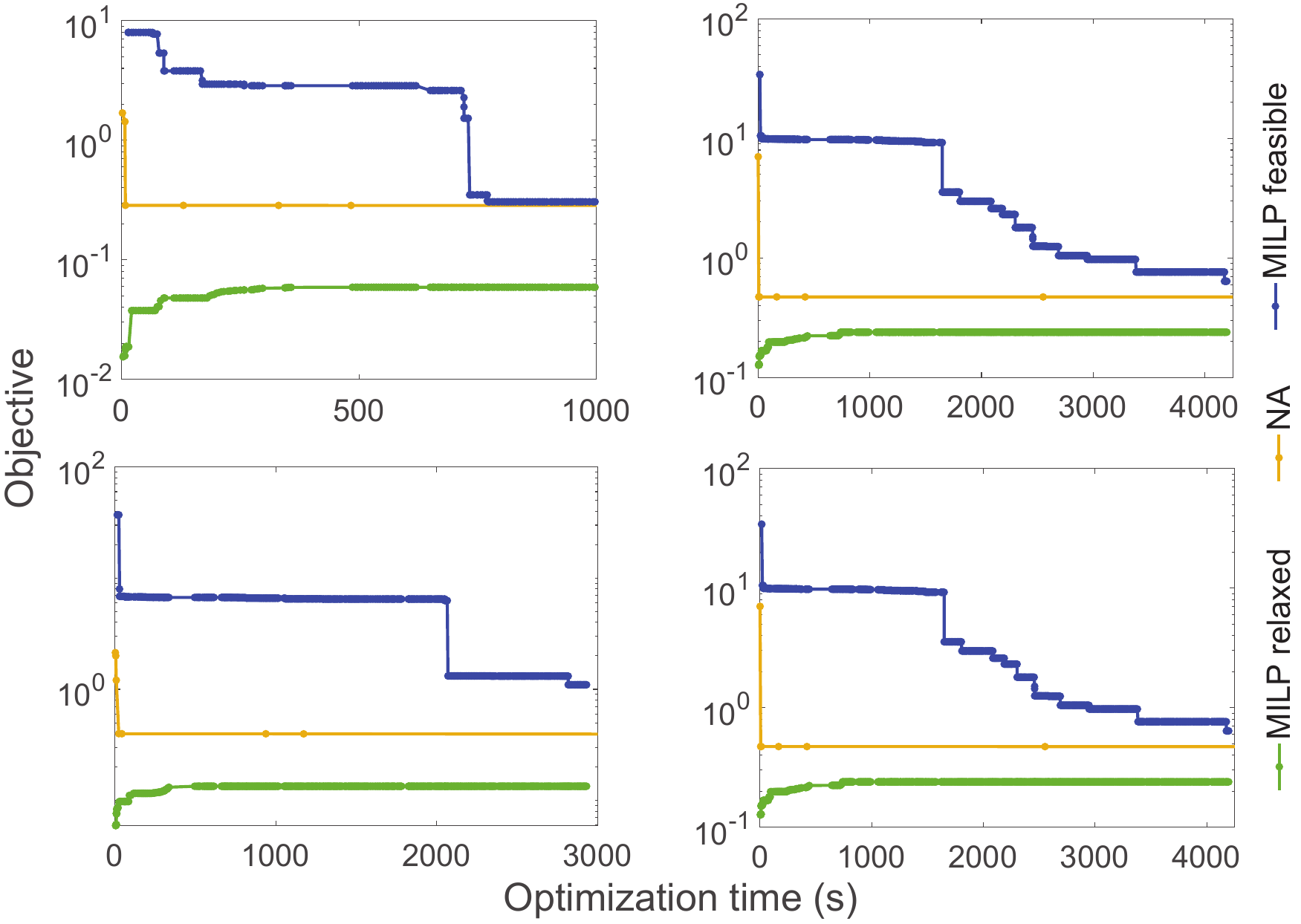}
          \caption{A NA-MILP hybrid approach. NA closes the optimality gap significantly faster than MILP's own objective's feasible solutions.} %
        \label{fig:NA_MILP_result}
\end{figure}

One of the greatest advantages of using MILP for inverse design is its optimality or near optimality guarantees. 
Despite this advantage, the MILP approach does not scale with larger networks. 
On the other hand, gradient-based local optimizers, such as NA \cite{NEURIPS2020_007ff380} are very efficient even for large networks. 
%
%
%
{As discussed in Section~\ref{sec:NaMilp}, NA solutions are feasible solutions to MILP's objective.
Therefore, we can produce feasible solutions via NA and continue using MILP for improving the relaxed solution.
}
In Figure~\ref{fig:NA_MILP_result}, we show this approach on 4 randomly target spectra in a spectral separation problem. 
We use a larger PL-NSM consisting of 4 layers of 150 neurons, mapping an 8D ink ratio input to spectra.
In all the experiments, NA improves {the feasible solution} significantly faster than MILP and by comparing NA solutions with the MILP's {relaxed solution} we can reach smaller optimality gaps ({distance between the yellow and the green lines instead of the distance between blue and green lines}) in a considerably less amount of time.
Although we use NA for computing {feasible solutions}, any other methods capable of yielding feasible solutions efficiently, could be used.
Finally, we would like to remind that this technique is only suitable for non-combinatorial inverse design problems where all solutions are trivially feasible.

\subsection{Robustness Analysis} \label{sec:robustnessexperiments}
\begin{figure}
     \centering
         \centering
         \includegraphics[width=0.5\textwidth]{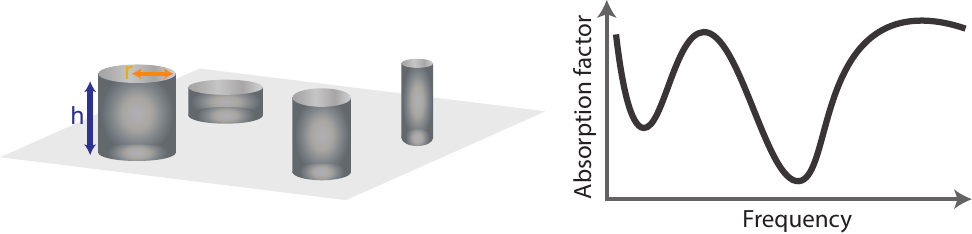}
          \caption{Schematic representation of an AMD metasurface. By adjusting the sub-wavelength nono-structures (cylinders' heights and radii), we can modulate the spectral absorption profile (right).} %
        \label{fig:AMD}
\end{figure}

\begin{figure}
     \centering
          \begin{subfigure}[b]{0.22\textwidth}
         \centering
         \includegraphics[width=\textwidth]{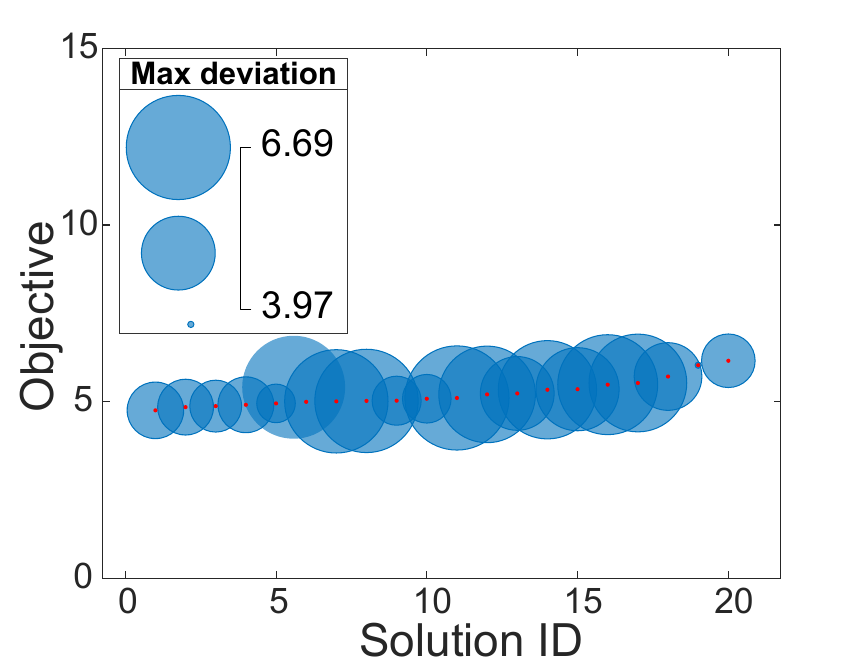}
         \caption{}
         \label{fig:roustness1}
     \end{subfigure}
\hfill             
    \begin{subfigure}[b]{0.22\textwidth}
         \centering
         \includegraphics[width=\textwidth]{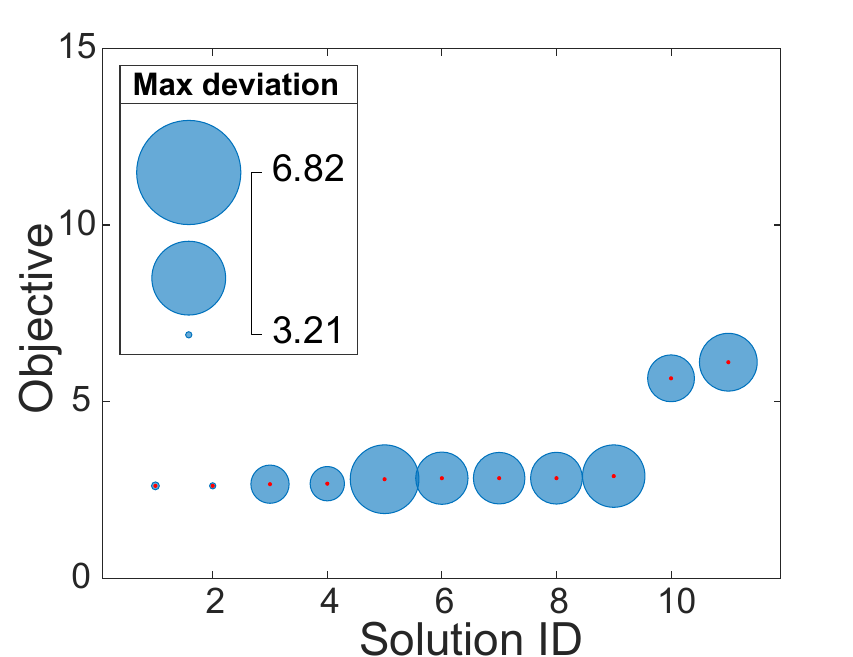}
         \caption{}
         \label{fig:roustness4}
     \end{subfigure}
\hfill    

        \caption{Robustness analysis of two target spectra of the metasurface design experiment. The size of the circle shows maximum deviation from the objective (Equation~\ref{eq:robustness}) and has an inverse relationship with robustness.} %
        \label{fig:robustness_analysis}
\end{figure}

For analyzing the robustness of different designs (Section~\ref{sec:robustnesstheory}), we use the recently proposed all-dielectric metasurface (ADM) setup \cite{NEURIPS2020_007ff380,nadell2019deep}. 
{ADMs are surfaces with nano-structures which possess unique properties like high temperature resistance, zero ohmic loss, and low thermal conductivity.}
{The properties of AMDs can be modulated by adjusting the structures and arrangement of these nano-structures.}
In this experiment, the spectral absorption of a particular AMD metasurface can be controlled via changing the heights and radii of nano-cylinders on its surface (Figure~\ref{fig:AMD}).
As there exist 4 nano-cylinders, the designs space can be expressed by 8 parameters. 
The resulting absorption spectrum is sampled at 300 points.
We train a convolutional PL-NSM that maps {8D} designs to {300D} spectra.
The network is made of 4 fully connected layers each with 500 neurons with ReLU activation functions and batch normalization, followed by 3 deconvolution and 1 convolutional layer. 
In our experiment, targeting two different spectra, we find a number of corresponding designs via NA which have the best objectives. 
In Figure~\ref{fig:robustness_analysis}, we show the (sorted) objectives and corresponding robustness for each solution. 
While, in Figure~\ref{fig:roustness1}, all objectives are comparable, there is one design (solution 19) that has a significantly higher robustness (indicated by a very small circle). 
This solution is the design of choice for this target spectrum. 
Moreover, in Figure~\ref{fig:roustness4}, sample 1 and 2 give very good accuracy and robustness at the same time and are clearly the most preferred designs. %
In this experiment, we spend on average 64 seconds on robustness calculation of each design and $3.64$ hours on a one-time bound precomputation. 
We also set $\epsilon$ (Equation~\ref{eq:robustness}) to $10^{-3}$. 
While here we use a simple perturbation model for evaluating the robustness, more sophisticated perturbations, such as erosion and dilation of designs \cite{sigmund2009manufacturing} are possible and left for future work.

\subsection{Scalability of MILP-Based Neural Inverse Design} \label{sec:milpnnanalysis}
\begin{figure}
     \centering
     \begin{subfigure}[b]{0.23\textwidth}
         \centering
         \includegraphics[width=\textwidth]{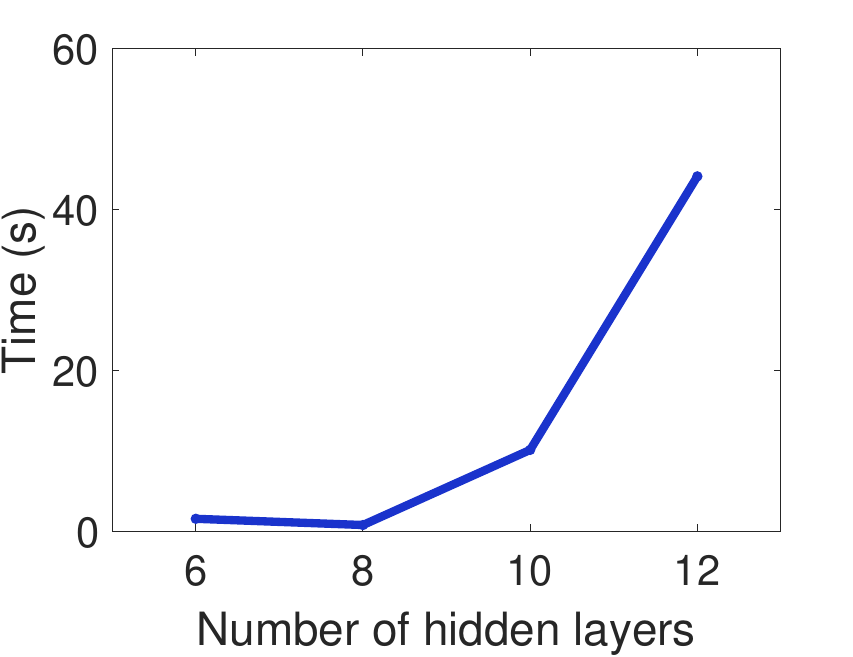}
         \caption{}
         \label{fig:depth_size}
     \end{subfigure}
\hfill     
          \begin{subfigure}[b]{0.23\textwidth}
         \centering
         \includegraphics[width=\textwidth]{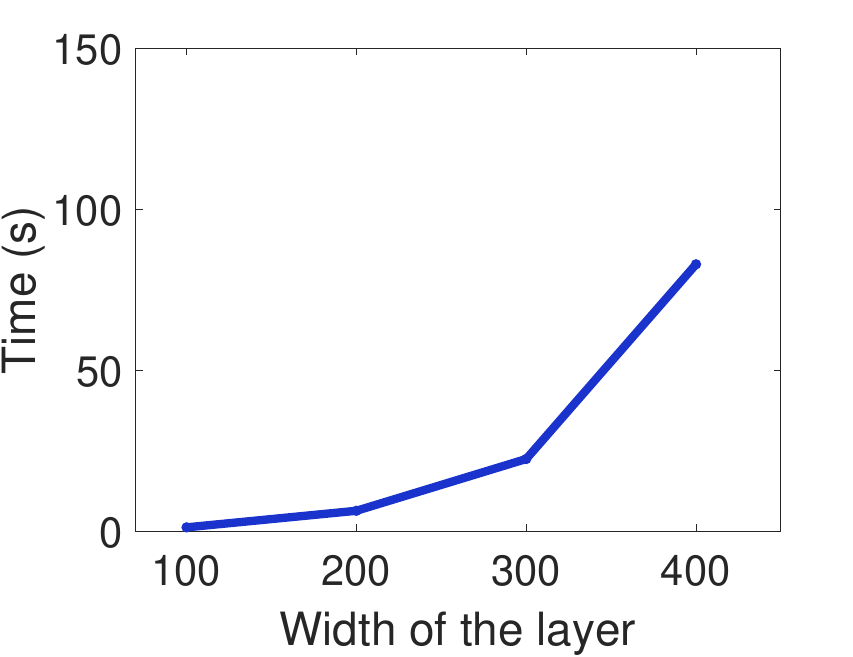}
         \caption{}
         \label{fig:width_size}
     \end{subfigure}
\hfill

        \caption{Scalability of our MILP-based neural inverse design with the depth and width of the network.}
        \label{fig:MILP_analysis}
\end{figure}

While MILPs are known to be NP-hard problems \cite{bunel2020branch}, it is interesting to study their scalability in our context. 
We choose the neural spectral separation experiment (Section~\ref{sec:resultinversedesign}) as a case study for our scalability analysis.  
First, in order to see the effect of network's depth on the solve time, we train 4 different PL-NSMs that perform forward spectral prediction for 8 inks. 
The trained networks have from 6 to 12 hidden layers, each with {50} neurons. 
In Figure~\ref{fig:depth_size}, we show the solve time for each of these networks averaged for {10} target spectra. 
Similarly, in Figure~\ref{fig:width_size}, we study the effect of network's width, by solving the same spectral separation problem performed on 4 PL-NSMs with a single hidden layer of 100 to 400 neurons. 
Here also the reported time is the average for {$10$} different target spectra. 
Once again, in all experiments, we continue the optimization to reach a duality gap of $0$ and thus global optima.

We observe that increasing the depth and width of the network increases the solve time, as expected in MILP problems. 
This is because each new node ({with unstable ReLU}) in the network gives rise to an additional binary and continuous variable, as well as new linear and integer constraints, in Equation~\ref{eq:relumilp}. 

\section{Limitations and Future Work} \label{sec:limitations}
Due to the NP-hardness of solving MILPs, this method is not suitable for large networks when searching for globally optimal designs. 
We nevertheless solved some real-life inverse design problems \cite{ansari2020mixed,shi2018deep,peurifoy2018nanophotonic} using this tool throughout this paper.
Looking at Figure~\ref{fig:depth_size}, we have found the globally optimal solution through inverting a neural network with 12 hidden layers each having 50 nodes in less than one minute. 
Thanks to the immense expressive power of neural networks \cite{hornik1989multilayer}, such an architecture is capable of accurately replacing many complex simulations. 
In fact, in this paper we trained forward models from the literature \cite{ansari2020mixed,shi2018deep} with smaller networks. 
In all these training, we used the same data with the same dimensionality of design and performance spaces and obtained nearly the same training error. 
In circumstances where using larger networks is necessary, the {relaxed solution} provided by the MILP solver help making informed decision on early stopping of the optimization. 
In such cases, the improvement of the {feasible solution} can also be accelerated via alternative solvers. 
An interesting direction for future work is to develop a solver customized to the type of inverse problems we deal with. 
We believe that neural networks with their recursive compositions are amenable to tailored heuristics beyond those found in one-size-fits-all solvers \cite{gurobi}. 
Ironically, machine learning is believed to be a potentially adept tool for discovering such heuristics \cite{NIPS2017_d9896106}. 
Finally, in addition to an accurate neural inversion method, the prerequisite for a perfect neural inverse design pipeline is an accurately trained NSM. 
While in our evaluations we ensure our trained NSMs are highly accurate, the focus of this work is on accurate optimization whose quality is measurable by the objective value of the optimization.  
A neural inversion method robust to training imperfections is highly interesting and is left for future work. 

\section{Conclusion} \label{sec:conclusion}
Neural networks are becoming first class citizens when it comes to data-driven modeling in computational design and fabrication. 
The black box reputation of neural networks has hindered applying them in domains requiring interpretability.
While this may to some extent be true during their training, once trained, neural networks are rather well-behaved mathematical functions.
In this work we showed that leveraging the underlying mathematical structure of neural surrogate models leads to a tool with many attractive properties. %
We believe our work paves the way for making sense of data-driven design processes in a more systematic manner. 
\appendix
\section{Bound precomputation algorithm}\label{sec:alg_appendix}
Extended bound precomputation algorithm calculates the upper and lower bound of each node in the neural network. 
All the upper and lower bounds in this algorithm are calculated under the constraints of the design and performance of our problem.
Considering these constraints results in tighter bounds and consequently more stable ReLU units, both of which make the inversion easier to calculate.

\begin{algorithm}
\SetAlgoLined

\textbf{Input} \\
$F_{\theta}$ \hspace{0.3cm} \textcolor{comment_color}{\tcp{Trained neural network}}
$t$\textsubscript{max}  \hspace{0.3cm} \textcolor{comment_color}{\tcp{Time limit for optimization}}
Design constraints \hspace{0.3cm} \textcolor{comment_color}{\tcp{e.g., fabrication constraints}}

\textbf{Output} \\
Constraints \hspace{0.32cm} \textcolor{comment_color}{\tcp{The set of all constraints including the upper and lower bounds}}

\Begin{
Constraints $\gets$ Design constraints \\
\For{$l\gets1$ \KwTo $L$} { \hspace{0.32cm} \textcolor{comment_color}{\tcp{Layers}}
        Optimizer $\gets$ Constraints \hspace{0.3cm} \textcolor{comment_color}{\tcp{ Updating the optimizer with new constraints after proceeding to the next layer}}
        
    \For{$k\gets1$ \KwTo $K$}{\hspace{0.32cm} \textcolor{comment_color}{\tcp{Nodes at layer $l$}}

        Start Timer\\
        Optimizer $\gets$ Obj (Equation \ref{eq:lbOptimization})\\
 
        \While{Optimizer}{
            \If{$Timer \geq$ $t$\textsubscript{max} $\algorithmicor  \  \ gap==0$}{
                $\mathbf{l}^{l}_{k} =  MILP_{relaxed}$  \hspace{0.3cm} \textcolor{comment_color}{\tcp{Relaxed solution determines the bound}}
                Constraints $\gets  \mathbf{l}^{l}_{k}$ \\
                Break
                \hspace{0.3cm} \textcolor{comment_color}{\tcp{Reaching the time limit or finding the optimal solution stops the optimization}}
              }
          }
        Start Timer \\
        Optimizer $\gets$ Obj (Equation \ref{eq:ubOptimization})\\
        \While{Optimizer}{
        \If{$Timer \geq$ $t$\textsubscript{max} $\algorithmicor  \  \ gap==0$}{

          $\mathbf{u}^{l}_{k} =  MILP_{relaxed}$  \\
          Constraints $\gets  \mathbf{u}^{l}_{k}$ 

            Break
            }
         }
     }

    }
}
\caption{Nodes' Lower and Upper Bound Precomputation.}
\label{alg:precomputation}
\end{algorithm}

\section{Soft robot inverse kinematics Objective for Neural adjoint Method}\label{sec:NA_softrobot_appendix}

We have also solved the inverse kinematics soft robot problem using the method of NA.
We use the objective function introduced in~\cite{sun2021amortized} with slight changes such that we can compare the errors with the results of the MILP method. 
The objective function is made of three terms:
\begin{equation}
\begin{split}
\mathcal{L}_{g}(\boldsymbol{\theta}, \boldsymbol{u}):=\left\| \mathbf{x}_{i}^{L} -\mathbf{t}\right\|_{1}+\lambda_{1} \cdot \mathcal{B}\left(\mathbf{x}^{L},  \mathbf{o}\right)+\lambda_{2} \cdot \mathcal{R}(\mathbf{x}^{0})\\
i \in \left[123, \ 124  \right],
\label{eq:SoftRobotObj_NA}
\end{split}
\end{equation}
\begin{equation}
\mathcal{B}\left(\mathbf{x}^{L},  \mathbf{o} \right):= \sum_{i=1}^{m}\left(\max \left(r \cdot \sqrt{2} +\Delta r-\left\|\mathbf{x}^{L}_{i}- \mathbf{o}\right\|_{1}, 0\right)\right)^{2},
\label{eq:SoftRobot_obtacle_NA}
\end{equation}
\begin{equation}
\mathcal{R}(\mathbf{x}^{0}):= \sum_{1<i<n, i \neq n / 2, \atop i \neq n / 2+1}
\left( \frac{\mathbf{x}^{0}_{i+1}-\mathbf{x}^{0}_{i}}{2}-\frac{\mathbf{x}^{0}_{i}-\mathbf{x}^{0}_{i-1}}{2}\right)^{2}.\\
\label{eq:SoftRobot_smoothness_NA}
\end{equation}
The $\mathcal{L}_{1}$ norm in Equation \ref{eq:SoftRobotObj_NA} is responsible for decreasing the distance between the tip of the soft robot and the target point.
Equation~\ref{eq:SoftRobot_obtacle_NA} punishes the possible collision of the soft robot with the obstacle and Equation~\ref{eq:SoftRobot_smoothness_NA} encourages the smoothness of the soft robot.
Finally, $\lambda_{1}$ and $\lambda_{2}$ are hyperparameters for balancing the objective, which we tune to $0.9$ and $0.1$, respectively.

\bibliographystyle{ACM-Reference-Format}
\bibliography{NSM_bib}


\begin{thebibliography}{48}


\ifx \showCODEN    \undefined \def \showCODEN     #1{\unskip}     \fi
\ifx \showDOI      \undefined \def \showDOI       #1{#1}\fi
\ifx \showISBNx    \undefined \def \showISBNx     #1{\unskip}     \fi
\ifx \showISBNxiii \undefined \def \showISBNxiii  #1{\unskip}     \fi
\ifx \showISSN     \undefined \def \showISSN      #1{\unskip}     \fi
\ifx \showLCCN     \undefined \def \showLCCN      #1{\unskip}     \fi
\ifx \shownote     \undefined \def \shownote      #1{#1}          \fi
\ifx \showarticletitle \undefined \def \showarticletitle #1{#1}   \fi
\ifx \showURL      \undefined \def \showURL       {\relax}        \fi
\providecommand\bibfield[2]{#2}
\providecommand\bibinfo[2]{#2}
\providecommand\natexlab[1]{#1}
\providecommand\showeprint[2][]{arXiv:#2}

\bibitem[\protect\citeauthoryear{Ansari, Alizadeh-Mousavi, Seidel, and
  Babaei}{Ansari et~al\mbox{.}}{2020}]%
        {ansari2020mixed}
\bibfield{author}{\bibinfo{person}{Navid Ansari}, \bibinfo{person}{Omid
  Alizadeh-Mousavi}, \bibinfo{person}{Hans-Peter Seidel}, {and}
  \bibinfo{person}{Vahid Babaei}.} \bibinfo{year}{2020}\natexlab{}.
\newblock \showarticletitle{Mixed integer ink selection for spectral
  reproduction}.
\newblock \bibinfo{journal}{\emph{ACM Transactions on Graphics (TOG)}}
  \bibinfo{volume}{39}, \bibinfo{number}{6} (\bibinfo{year}{2020}),
  \bibinfo{pages}{1--16}.
\newblock


\bibitem[\protect\citeauthoryear{Ardizzone, Kruse, Rother, and
  Köthe}{Ardizzone et~al\mbox{.}}{2019}]%
        {ardizzone2018analyzing}
\bibfield{author}{\bibinfo{person}{Lynton Ardizzone}, \bibinfo{person}{Jakob
  Kruse}, \bibinfo{person}{Carsten Rother}, {and} \bibinfo{person}{Ullrich
  Köthe}.} \bibinfo{year}{2019}\natexlab{}.
\newblock \showarticletitle{Analyzing Inverse Problems with Invertible Neural
  Networks}. In \bibinfo{booktitle}{\emph{International Conference on Learning
  Representations}}.
\newblock
\urldef\tempurl%
\url{https://openreview.net/forum?id=rJed6j0cKX}
\showURL{%
\tempurl}


\bibitem[\protect\citeauthoryear{Babaei, Vidim{\v{c}}e, Foshey, Kaspar, Didyk,
  and Matusik}{Babaei et~al\mbox{.}}{2017}]%
        {babaei2017color}
\bibfield{author}{\bibinfo{person}{Vahid Babaei}, \bibinfo{person}{Kiril
  Vidim{\v{c}}e}, \bibinfo{person}{Michael Foshey}, \bibinfo{person}{Alexandre
  Kaspar}, \bibinfo{person}{Piotr Didyk}, {and} \bibinfo{person}{Wojciech
  Matusik}.} \bibinfo{year}{2017}\natexlab{}.
\newblock \showarticletitle{Color contoning for 3D printing}.
\newblock \bibinfo{journal}{\emph{ACM Transactions on Graphics (TOG)}}
  \bibinfo{volume}{36}, \bibinfo{number}{4} (\bibinfo{year}{2017}),
  \bibinfo{pages}{1--15}.
\newblock


\bibitem[\protect\citeauthoryear{Belotti, Kirches, Leyffer, Linderoth, Luedtke,
  and Mahajan}{Belotti et~al\mbox{.}}{2013}]%
        {belotti2013mixed}
\bibfield{author}{\bibinfo{person}{Pietro Belotti}, \bibinfo{person}{Christian
  Kirches}, \bibinfo{person}{Sven Leyffer}, \bibinfo{person}{Jeff Linderoth},
  \bibinfo{person}{James Luedtke}, {and} \bibinfo{person}{Ashutosh Mahajan}.}
  \bibinfo{year}{2013}\natexlab{}.
\newblock \showarticletitle{Mixed-integer nonlinear optimization}.
\newblock \bibinfo{journal}{\emph{Acta Numerica}}  \bibinfo{volume}{22}
  (\bibinfo{year}{2013}), \bibinfo{pages}{1--131}.
\newblock


\bibitem[\protect\citeauthoryear{Bermano, Funkhouser, and Rusinkiewicz}{Bermano
  et~al\mbox{.}}{2017}]%
        {bermano2017state}
\bibfield{author}{\bibinfo{person}{Amit~H Bermano}, \bibinfo{person}{Thomas
  Funkhouser}, {and} \bibinfo{person}{Szymon Rusinkiewicz}.}
  \bibinfo{year}{2017}\natexlab{}.
\newblock \showarticletitle{State of the art in methods and representations for
  fabrication-aware design}. In \bibinfo{booktitle}{\emph{Computer Graphics
  Forum}}, Vol.~\bibinfo{volume}{36}. Wiley Online Library,
  \bibinfo{pages}{509--535}.
\newblock


\bibitem[\protect\citeauthoryear{Bertoldi, Vitelli, Christensen, and
  Van~Hecke}{Bertoldi et~al\mbox{.}}{2017}]%
        {bertoldi2017flexible}
\bibfield{author}{\bibinfo{person}{Katia Bertoldi}, \bibinfo{person}{Vincenzo
  Vitelli}, \bibinfo{person}{Johan Christensen}, {and} \bibinfo{person}{Martin
  Van~Hecke}.} \bibinfo{year}{2017}\natexlab{}.
\newblock \showarticletitle{Flexible mechanical metamaterials}.
\newblock \bibinfo{journal}{\emph{Nature Reviews Materials}}
  \bibinfo{volume}{2}, \bibinfo{number}{11} (\bibinfo{year}{2017}),
  \bibinfo{pages}{1--11}.
\newblock


\bibitem[\protect\citeauthoryear{Bunel, Lu, Turkaslan, Kohli, Torr, and
  Mudigonda}{Bunel et~al\mbox{.}}{2020}]%
        {bunel2020branch}
\bibfield{author}{\bibinfo{person}{Rudy Bunel}, \bibinfo{person}{Jingyue Lu},
  \bibinfo{person}{Ilker Turkaslan}, \bibinfo{person}{Pushmeet Kohli},
  \bibinfo{person}{P Torr}, {and} \bibinfo{person}{P Mudigonda}.}
  \bibinfo{year}{2020}\natexlab{}.
\newblock \showarticletitle{Branch and bound for piecewise linear neural
  network verification}.
\newblock \bibinfo{journal}{\emph{Journal of Machine Learning Research}}
  \bibinfo{volume}{21}, \bibinfo{number}{2020} (\bibinfo{year}{2020}).
\newblock


\bibitem[\protect\citeauthoryear{Bunel, Turkaslan, Torr, Kohli, and
  Mudigonda}{Bunel et~al\mbox{.}}{2018}]%
        {NEURIPS2018_be53d253}
\bibfield{author}{\bibinfo{person}{Rudy~R Bunel}, \bibinfo{person}{Ilker
  Turkaslan}, \bibinfo{person}{Philip Torr}, \bibinfo{person}{Pushmeet Kohli},
  {and} \bibinfo{person}{Pawan~K Mudigonda}.} \bibinfo{year}{2018}\natexlab{}.
\newblock \showarticletitle{A Unified View of Piecewise Linear Neural Network
  Verification}. In \bibinfo{booktitle}{\emph{Advances in Neural Information
  Processing Systems}}, \bibfield{editor}{\bibinfo{person}{S.~Bengio},
  \bibinfo{person}{H.~Wallach}, \bibinfo{person}{H.~Larochelle},
  \bibinfo{person}{K.~Grauman}, \bibinfo{person}{N.~Cesa-Bianchi}, {and}
  \bibinfo{person}{R.~Garnett}} (Eds.), Vol.~\bibinfo{volume}{31}.
  \bibinfo{publisher}{Curran Associates, Inc.}, \bibinfo{pages}{4790--4799}.
\newblock
\urldef\tempurl%
\url{https://proceedings.neurips.cc/paper/2018/file/be53d253d6bc3258a8160556dda3e9b2-Paper.pdf}
\showURL{%
\tempurl}


\bibitem[\protect\citeauthoryear{Chen, Levin, Didyk, Sitthi-Amorn, and
  Matusik}{Chen et~al\mbox{.}}{2013}]%
        {chen2013spec2fab}
\bibfield{author}{\bibinfo{person}{Desai Chen}, \bibinfo{person}{David~IW
  Levin}, \bibinfo{person}{Piotr Didyk}, \bibinfo{person}{Pitchaya
  Sitthi-Amorn}, {and} \bibinfo{person}{Wojciech Matusik}.}
  \bibinfo{year}{2013}\natexlab{}.
\newblock \showarticletitle{Spec2Fab: a reducer-tuner model for translating
  specifications to 3D prints}.
\newblock \bibinfo{journal}{\emph{ACM Transactions on Graphics (TOG)}}
  \bibinfo{volume}{32}, \bibinfo{number}{4} (\bibinfo{year}{2013}),
  \bibinfo{pages}{135}.
\newblock


\bibitem[\protect\citeauthoryear{Cheng, N{\"u}hrenberg, and Ruess}{Cheng
  et~al\mbox{.}}{2017}]%
        {cheng2017maximum}
\bibfield{author}{\bibinfo{person}{Chih-Hong Cheng}, \bibinfo{person}{Georg
  N{\"u}hrenberg}, {and} \bibinfo{person}{Harald Ruess}.}
  \bibinfo{year}{2017}\natexlab{}.
\newblock \showarticletitle{Maximum resilience of artificial neural networks}.
  In \bibinfo{booktitle}{\emph{International Symposium on Automated Technology
  for Verification and Analysis}}. Springer, \bibinfo{pages}{251--268}.
\newblock


\bibitem[\protect\citeauthoryear{Dinh, Sohl{-}Dickstein, and Bengio}{Dinh
  et~al\mbox{.}}{2017}]%
        {DBLP:conf/iclr/DinhSB17}
\bibfield{author}{\bibinfo{person}{Laurent Dinh}, \bibinfo{person}{Jascha
  Sohl{-}Dickstein}, {and} \bibinfo{person}{Samy Bengio}.}
  \bibinfo{year}{2017}\natexlab{}.
\newblock \showarticletitle{Density estimation using Real {NVP}}. In
  \bibinfo{booktitle}{\emph{5th International Conference on Learning
  Representations, {ICLR} 2017, Toulon, France, April 24-26, 2017, Conference
  Track Proceedings}}. \bibinfo{publisher}{OpenReview.net}.
\newblock
\urldef\tempurl%
\url{https://openreview.net/forum?id=HkpbnH9lx}
\showURL{%
\tempurl}


\bibitem[\protect\citeauthoryear{Fischetti and Jo}{Fischetti and Jo}{2018}]%
        {fischetti2018deep}
\bibfield{author}{\bibinfo{person}{Matteo Fischetti} {and}
  \bibinfo{person}{Jason Jo}.} \bibinfo{year}{2018}\natexlab{}.
\newblock \showarticletitle{Deep neural networks and mixed integer linear
  optimization}.
\newblock \bibinfo{journal}{\emph{Constraints}} \bibinfo{volume}{23},
  \bibinfo{number}{3} (\bibinfo{year}{2018}), \bibinfo{pages}{296--309}.
\newblock


\bibitem[\protect\citeauthoryear{Floudas}{Floudas}{1995}]%
        {floudas1995nonlinear}
\bibfield{author}{\bibinfo{person}{Christodoulos~A Floudas}.}
  \bibinfo{year}{1995}\natexlab{}.
\newblock \bibinfo{booktitle}{\emph{Nonlinear and mixed-integer optimization:
  fundamentals and applications}}.
\newblock \bibinfo{publisher}{Oxford University Press}.
\newblock


\bibitem[\protect\citeauthoryear{Gavriil, Guseinov, P{\'e}rez, Pellis,
  Henderson, Rist, Pottmann, and Bickel}{Gavriil et~al\mbox{.}}{2020}]%
        {gavriil2020computational}
\bibfield{author}{\bibinfo{person}{Konstantinos Gavriil},
  \bibinfo{person}{Ruslan Guseinov}, \bibinfo{person}{Jes{\'u}s P{\'e}rez},
  \bibinfo{person}{Davide Pellis}, \bibinfo{person}{Paul Henderson},
  \bibinfo{person}{Florian Rist}, \bibinfo{person}{Helmut Pottmann}, {and}
  \bibinfo{person}{Bernd Bickel}.} \bibinfo{year}{2020}\natexlab{}.
\newblock \showarticletitle{Computational design of cold bent glass
  fa{\c{c}}ades}.
\newblock \bibinfo{journal}{\emph{ACM Transactions on Graphics (TOG)}}
  \bibinfo{volume}{39}, \bibinfo{number}{6} (\bibinfo{year}{2020}),
  \bibinfo{pages}{1--16}.
\newblock


\bibitem[\protect\citeauthoryear{Gurobi~Optimization}{Gurobi~Optimization}{2018}]%
        {gurobi}
\bibfield{author}{\bibinfo{person}{LLC Gurobi~Optimization}.}
  \bibinfo{year}{2018}\natexlab{}.
\newblock \bibinfo{title}{Gurobi Optimizer Reference Manual}.
\newblock
\newblock
\urldef\tempurl%
\url{http://www.gurobi.com}
\showURL{%
\tempurl}


\bibitem[\protect\citeauthoryear{Hecht-Nielsen}{Hecht-Nielsen}{1992}]%
        {hecht1992theory}
\bibfield{author}{\bibinfo{person}{Robert Hecht-Nielsen}.}
  \bibinfo{year}{1992}\natexlab{}.
\newblock \showarticletitle{Theory of the backpropagation neural network}.
\newblock In \bibinfo{booktitle}{\emph{Neural networks for perception}}.
  \bibinfo{publisher}{Elsevier}, \bibinfo{pages}{65--93}.
\newblock


\bibitem[\protect\citeauthoryear{Hornik, Stinchcombe, and White}{Hornik
  et~al\mbox{.}}{1989}]%
        {hornik1989multilayer}
\bibfield{author}{\bibinfo{person}{Kurt Hornik}, \bibinfo{person}{Maxwell
  Stinchcombe}, {and} \bibinfo{person}{Halbert White}.}
  \bibinfo{year}{1989}\natexlab{}.
\newblock \showarticletitle{Multilayer feedforward networks are universal
  approximators}.
\newblock \bibinfo{journal}{\emph{Neural networks}} \bibinfo{volume}{2},
  \bibinfo{number}{5} (\bibinfo{year}{1989}), \bibinfo{pages}{359--366}.
\newblock


\bibitem[\protect\citeauthoryear{Jiang, Chen, and Fan}{Jiang
  et~al\mbox{.}}{2020}]%
        {jiang2020deep}
\bibfield{author}{\bibinfo{person}{Jiaqi Jiang}, \bibinfo{person}{Mingkun
  Chen}, {and} \bibinfo{person}{Jonathan~A Fan}.}
  \bibinfo{year}{2020}\natexlab{}.
\newblock \showarticletitle{Deep neural networks for the evaluation and design
  of photonic devices}.
\newblock \bibinfo{journal}{\emph{Nature Reviews Materials}}
  (\bibinfo{year}{2020}), \bibinfo{pages}{1--22}.
\newblock


\bibitem[\protect\citeauthoryear{Khalil, Dai, Zhang, Dilkina, and Song}{Khalil
  et~al\mbox{.}}{2017}]%
        {NIPS2017_d9896106}
\bibfield{author}{\bibinfo{person}{Elias Khalil}, \bibinfo{person}{Hanjun Dai},
  \bibinfo{person}{Yuyu Zhang}, \bibinfo{person}{Bistra Dilkina}, {and}
  \bibinfo{person}{Le Song}.} \bibinfo{year}{2017}\natexlab{}.
\newblock \showarticletitle{Learning Combinatorial Optimization Algorithms over
  Graphs}. In \bibinfo{booktitle}{\emph{Advances in Neural Information
  Processing Systems}}, \bibfield{editor}{\bibinfo{person}{I.~Guyon},
  \bibinfo{person}{U.~V. Luxburg}, \bibinfo{person}{S.~Bengio},
  \bibinfo{person}{H.~Wallach}, \bibinfo{person}{R.~Fergus},
  \bibinfo{person}{S.~Vishwanathan}, {and} \bibinfo{person}{R.~Garnett}}
  (Eds.), Vol.~\bibinfo{volume}{30}. \bibinfo{publisher}{Curran Associates,
  Inc.}
\newblock
\urldef\tempurl%
\url{https://proceedings.neurips.cc/paper/2017/file/d9896106ca98d3d05b8cbdf4fd8b13a1-Paper.pdf}
\showURL{%
\tempurl}


\bibitem[\protect\citeauthoryear{Kiarashinejad, Abdollahramezani, and
  Adibi}{Kiarashinejad et~al\mbox{.}}{2020}]%
        {kiarashinejad2020deep}
\bibfield{author}{\bibinfo{person}{Yashar Kiarashinejad},
  \bibinfo{person}{Sajjad Abdollahramezani}, {and} \bibinfo{person}{Ali
  Adibi}.} \bibinfo{year}{2020}\natexlab{}.
\newblock \showarticletitle{Deep learning approach based on dimensionality
  reduction for designing electromagnetic nanostructures}.
\newblock \bibinfo{journal}{\emph{npj Computational Materials}}
  \bibinfo{volume}{6}, \bibinfo{number}{1} (\bibinfo{year}{2020}),
  \bibinfo{pages}{1--12}.
\newblock


\bibitem[\protect\citeauthoryear{Kingma and Ba}{Kingma and Ba}{2014}]%
        {kingma2014adam}
\bibfield{author}{\bibinfo{person}{Diederik~P Kingma} {and}
  \bibinfo{person}{Jimmy Ba}.} \bibinfo{year}{2014}\natexlab{}.
\newblock \showarticletitle{Adam: A method for stochastic optimization}.
\newblock \bibinfo{journal}{\emph{arXiv preprint arXiv:1412.6980}}
  (\bibinfo{year}{2014}).
\newblock


\bibitem[\protect\citeauthoryear{Kingma and Welling}{Kingma and
  Welling}{2013}]%
        {kingma2013auto}
\bibfield{author}{\bibinfo{person}{Diederik~P Kingma} {and}
  \bibinfo{person}{Max Welling}.} \bibinfo{year}{2013}\natexlab{}.
\newblock \showarticletitle{Auto-encoding variational bayes}.
\newblock \bibinfo{journal}{\emph{arXiv preprint arXiv:1312.6114}}
  (\bibinfo{year}{2013}).
\newblock


\bibitem[\protect\citeauthoryear{Klotz and Newman}{Klotz and Newman}{2013}]%
        {klotz2013practical}
\bibfield{author}{\bibinfo{person}{Ed Klotz} {and} \bibinfo{person}{Alexandra~M
  Newman}.} \bibinfo{year}{2013}\natexlab{}.
\newblock \showarticletitle{Practical guidelines for solving difficult mixed
  integer linear programs}.
\newblock \bibinfo{journal}{\emph{Surveys in Operations Research and Management
  Science}} \bibinfo{volume}{18}, \bibinfo{number}{1-2} (\bibinfo{year}{2013}),
  \bibinfo{pages}{18--32}.
\newblock


\bibitem[\protect\citeauthoryear{Liu, Tan, Khoram, and Yu}{Liu
  et~al\mbox{.}}{2018}]%
        {liu2018training}
\bibfield{author}{\bibinfo{person}{Dianjing Liu}, \bibinfo{person}{Yixuan Tan},
  \bibinfo{person}{Erfan Khoram}, {and} \bibinfo{person}{Zongfu Yu}.}
  \bibinfo{year}{2018}\natexlab{}.
\newblock \showarticletitle{Training deep neural networks for the inverse
  design of nanophotonic structures}.
\newblock \bibinfo{journal}{\emph{ACS Photonics}} \bibinfo{volume}{5},
  \bibinfo{number}{4} (\bibinfo{year}{2018}), \bibinfo{pages}{1365--1369}.
\newblock


\bibitem[\protect\citeauthoryear{Majidi}{Majidi}{2014}]%
        {majidi2014soft}
\bibfield{author}{\bibinfo{person}{Carmel Majidi}.}
  \bibinfo{year}{2014}\natexlab{}.
\newblock \showarticletitle{Soft robotics: a perspective—current trends and
  prospects for the future}.
\newblock \bibinfo{journal}{\emph{Soft robotics}} \bibinfo{volume}{1},
  \bibinfo{number}{1} (\bibinfo{year}{2014}), \bibinfo{pages}{5--11}.
\newblock


\bibitem[\protect\citeauthoryear{Markowitz and Manne}{Markowitz and
  Manne}{1957}]%
        {markowitz1957solution}
\bibfield{author}{\bibinfo{person}{Harry~M Markowitz} {and}
  \bibinfo{person}{Alan~S Manne}.} \bibinfo{year}{1957}\natexlab{}.
\newblock \showarticletitle{On the solution of discrete programming problems}.
\newblock \bibinfo{journal}{\emph{Econometrica: journal of the Econometric
  Society}} (\bibinfo{year}{1957}), \bibinfo{pages}{84--110}.
\newblock


\bibitem[\protect\citeauthoryear{Matusik, Ajdin, Gu, Lawrence, Lensch,
  Pellacini, and Rusinkiewicz}{Matusik et~al\mbox{.}}{2009}]%
        {Matusik2009}
\bibfield{author}{\bibinfo{person}{Wojciech Matusik}, \bibinfo{person}{Boris
  Ajdin}, \bibinfo{person}{Jinwei Gu}, \bibinfo{person}{Jason Lawrence},
  \bibinfo{person}{Hendrik P.~A. Lensch}, \bibinfo{person}{Fabio Pellacini},
  {and} \bibinfo{person}{Szymon Rusinkiewicz}.}
  \bibinfo{year}{2009}\natexlab{}.
\newblock \showarticletitle{Printing Spatially-varying Reflectance}.
\newblock \bibinfo{journal}{\emph{ACM Trans. Graph.}} \bibinfo{volume}{28},
  \bibinfo{number}{5} (\bibinfo{date}{Dec.} \bibinfo{year}{2009}),
  \bibinfo{pages}{128:1--128:9}.
\newblock


\bibitem[\protect\citeauthoryear{Mitra and Pauly}{Mitra and Pauly}{2009}]%
        {mitra2009shadow}
\bibfield{author}{\bibinfo{person}{Niloy~J Mitra} {and} \bibinfo{person}{Mark
  Pauly}.} \bibinfo{year}{2009}\natexlab{}.
\newblock \showarticletitle{Shadow art}.
\newblock \bibinfo{journal}{\emph{ACM Transactions on Graphics}}
  \bibinfo{volume}{28}, \bibinfo{number}{CONF} (\bibinfo{year}{2009}),
  \bibinfo{pages}{156--1}.
\newblock


\bibitem[\protect\citeauthoryear{Montufar, Pascanu, Cho, and Bengio}{Montufar
  et~al\mbox{.}}{2014}]%
        {montufar2014number}
\bibfield{author}{\bibinfo{person}{Guido~F Montufar}, \bibinfo{person}{Razvan
  Pascanu}, \bibinfo{person}{Kyunghyun Cho}, {and} \bibinfo{person}{Yoshua
  Bengio}.} \bibinfo{year}{2014}\natexlab{}.
\newblock \showarticletitle{On the number of linear regions of deep neural
  networks}. In \bibinfo{booktitle}{\emph{Advances in neural information
  processing systems}}. \bibinfo{pages}{2924--2932}.
\newblock


\bibitem[\protect\citeauthoryear{Nadell, Huang, Malof, and Padilla}{Nadell
  et~al\mbox{.}}{2019}]%
        {nadell2019deep}
\bibfield{author}{\bibinfo{person}{Christian~C Nadell}, \bibinfo{person}{Bohao
  Huang}, \bibinfo{person}{Jordan~M Malof}, {and} \bibinfo{person}{Willie~J
  Padilla}.} \bibinfo{year}{2019}\natexlab{}.
\newblock \showarticletitle{Deep learning for accelerated all-dielectric
  metasurface design}.
\newblock \bibinfo{journal}{\emph{Optics express}} \bibinfo{volume}{27},
  \bibinfo{number}{20} (\bibinfo{year}{2019}), \bibinfo{pages}{27523--27535}.
\newblock


\bibitem[\protect\citeauthoryear{Nindel, Iser, Rittig, Wilkie, and
  Křivánek}{Nindel et~al\mbox{.}}{2021}]%
        {nindel2021gradient}
\bibfield{author}{\bibinfo{person}{Thomas Nindel}, \bibinfo{person}{Tomáš
  Iser}, \bibinfo{person}{Tobias Rittig}, \bibinfo{person}{Alexander Wilkie},
  {and} \bibinfo{person}{Jaroslav Křivánek}.}
  \bibinfo{year}{2021}\natexlab{}.
\newblock \showarticletitle{A Gradient-Based Framework for 3D Print Appearance
  Optimization}.
\newblock \bibinfo{journal}{\emph{ACM Transactions on Graphics (TOG)}}
  \bibinfo{volume}{40}, \bibinfo{number}{4} (\bibinfo{year}{2021}).
\newblock


\bibitem[\protect\citeauthoryear{Peurifoy, Shen, Jing, Yang, Cano-Renteria,
  DeLacy, Joannopoulos, Tegmark, and Solja{\v{c}}i{\'c}}{Peurifoy
  et~al\mbox{.}}{2018}]%
        {peurifoy2018nanophotonic}
\bibfield{author}{\bibinfo{person}{John Peurifoy}, \bibinfo{person}{Yichen
  Shen}, \bibinfo{person}{Li Jing}, \bibinfo{person}{Yi Yang},
  \bibinfo{person}{Fidel Cano-Renteria}, \bibinfo{person}{Brendan~G DeLacy},
  \bibinfo{person}{John~D Joannopoulos}, \bibinfo{person}{Max Tegmark}, {and}
  \bibinfo{person}{Marin Solja{\v{c}}i{\'c}}.} \bibinfo{year}{2018}\natexlab{}.
\newblock \showarticletitle{Nanophotonic particle simulation and inverse design
  using artificial neural networks}.
\newblock \bibinfo{journal}{\emph{Science advances}} \bibinfo{volume}{4},
  \bibinfo{number}{6} (\bibinfo{year}{2018}), \bibinfo{pages}{eaar4206}.
\newblock


\bibitem[\protect\citeauthoryear{Piovar{\v{c}}i, Foshey, Babaei, Rusinkiewicz,
  Matusik, and Didyk}{Piovar{\v{c}}i et~al\mbox{.}}{2020}]%
        {piovarvci2020towards}
\bibfield{author}{\bibinfo{person}{Michal Piovar{\v{c}}i},
  \bibinfo{person}{Michael Foshey}, \bibinfo{person}{Vahid Babaei},
  \bibinfo{person}{Szymon Rusinkiewicz}, \bibinfo{person}{Wojciech Matusik},
  {and} \bibinfo{person}{Piotr Didyk}.} \bibinfo{year}{2020}\natexlab{}.
\newblock \showarticletitle{Towards spatially varying gloss reproduction for 3D
  printing}.
\newblock \bibinfo{journal}{\emph{ACM Transactions on Graphics (TOG)}}
  \bibinfo{volume}{39}, \bibinfo{number}{6} (\bibinfo{year}{2020}),
  \bibinfo{pages}{1--13}.
\newblock


\bibitem[\protect\citeauthoryear{Polygerinos, Wang, Galloway, Wood, and
  Walsh}{Polygerinos et~al\mbox{.}}{2015}]%
        {polygerinos2015soft}
\bibfield{author}{\bibinfo{person}{Panagiotis Polygerinos},
  \bibinfo{person}{Zheng Wang}, \bibinfo{person}{Kevin~C Galloway},
  \bibinfo{person}{Robert~J Wood}, {and} \bibinfo{person}{Conor~J Walsh}.}
  \bibinfo{year}{2015}\natexlab{}.
\newblock \showarticletitle{Soft robotic glove for combined assistance and
  at-home rehabilitation}.
\newblock \bibinfo{journal}{\emph{Robotics and Autonomous Systems}}
  \bibinfo{volume}{73} (\bibinfo{year}{2015}), \bibinfo{pages}{135--143}.
\newblock


\bibitem[\protect\citeauthoryear{Ren, Padilla, and Malof}{Ren
  et~al\mbox{.}}{2020}]%
        {NEURIPS2020_007ff380}
\bibfield{author}{\bibinfo{person}{Simiao Ren}, \bibinfo{person}{Willie
  Padilla}, {and} \bibinfo{person}{Jordan Malof}.}
  \bibinfo{year}{2020}\natexlab{}.
\newblock \showarticletitle{Benchmarking Deep Inverse Models over time, and the
  Neural-Adjoint method}. In \bibinfo{booktitle}{\emph{Advances in Neural
  Information Processing Systems}},
  \bibfield{editor}{\bibinfo{person}{H.~Larochelle},
  \bibinfo{person}{M.~Ranzato}, \bibinfo{person}{R.~Hadsell},
  \bibinfo{person}{M.~F. Balcan}, {and} \bibinfo{person}{H.~Lin}} (Eds.),
  Vol.~\bibinfo{volume}{33}. \bibinfo{publisher}{Curran Associates, Inc.},
  \bibinfo{pages}{38--48}.
\newblock
\urldef\tempurl%
\url{https://proceedings.neurips.cc/paper/2020/file/007ff380ee5ac49ffc34442f5c2a2b86-Paper.pdf}
\showURL{%
\tempurl}


\bibitem[\protect\citeauthoryear{Sch{\"u}ller, Panozzo, and
  Sorkine-Hornung}{Sch{\"u}ller et~al\mbox{.}}{2014}]%
        {schuller2014appearance}
\bibfield{author}{\bibinfo{person}{Christian Sch{\"u}ller},
  \bibinfo{person}{Daniele Panozzo}, {and} \bibinfo{person}{Olga
  Sorkine-Hornung}.} \bibinfo{year}{2014}\natexlab{}.
\newblock \showarticletitle{Appearance-mimicking surfaces}.
\newblock \bibinfo{journal}{\emph{ACM Transactions on Graphics (TOG)}}
  \bibinfo{volume}{33}, \bibinfo{number}{6} (\bibinfo{year}{2014}),
  \bibinfo{pages}{1--10}.
\newblock


\bibitem[\protect\citeauthoryear{Schumacher, Bickel, Rys, Marschner, Daraio,
  and Gross}{Schumacher et~al\mbox{.}}{2015}]%
        {schumacher2015microstructures}
\bibfield{author}{\bibinfo{person}{Christian Schumacher},
  \bibinfo{person}{Bernd Bickel}, \bibinfo{person}{Jan Rys},
  \bibinfo{person}{Steve Marschner}, \bibinfo{person}{Chiara Daraio}, {and}
  \bibinfo{person}{Markus Gross}.} \bibinfo{year}{2015}\natexlab{}.
\newblock \showarticletitle{Microstructures to control elasticity in 3D
  printing}.
\newblock \bibinfo{journal}{\emph{ACM Transactions on Graphics (TOG)}}
  \bibinfo{volume}{34}, \bibinfo{number}{4} (\bibinfo{year}{2015}),
  \bibinfo{pages}{136}.
\newblock


\bibitem[\protect\citeauthoryear{Schwartzburg, Testuz, Tagliasacchi, and
  Pauly}{Schwartzburg et~al\mbox{.}}{2014}]%
        {schwartzburg2014high}
\bibfield{author}{\bibinfo{person}{Yuliy Schwartzburg}, \bibinfo{person}{Romain
  Testuz}, \bibinfo{person}{Andrea Tagliasacchi}, {and} \bibinfo{person}{Mark
  Pauly}.} \bibinfo{year}{2014}\natexlab{}.
\newblock \showarticletitle{High-contrast computational caustic design}.
\newblock \bibinfo{journal}{\emph{ACM Transactions on Graphics (TOG)}}
  \bibinfo{volume}{33}, \bibinfo{number}{4} (\bibinfo{year}{2014}),
  \bibinfo{pages}{1--11}.
\newblock


\bibitem[\protect\citeauthoryear{Shi, Babaei, Kim, Foshey, Hu, Sitthi-Amorn,
  Rusinkiewicz, and Matusik}{Shi et~al\mbox{.}}{2018}]%
        {shi2018deep}
\bibfield{author}{\bibinfo{person}{Liang Shi}, \bibinfo{person}{Vahid Babaei},
  \bibinfo{person}{Changil Kim}, \bibinfo{person}{Michael Foshey},
  \bibinfo{person}{Yuanming Hu}, \bibinfo{person}{Pitchaya Sitthi-Amorn},
  \bibinfo{person}{Szymon Rusinkiewicz}, {and} \bibinfo{person}{Wojciech
  Matusik}.} \bibinfo{year}{2018}\natexlab{}.
\newblock \showarticletitle{Deep multispectral painting reproduction via
  multi-layer, custom-ink printing}.
\newblock \bibinfo{journal}{\emph{ACM Trans. Graph.}} \bibinfo{volume}{37},
  \bibinfo{number}{6} (\bibinfo{date}{Dec.} \bibinfo{year}{2018}),
  \bibinfo{pages}{271:1--271:15}.
\newblock


\bibitem[\protect\citeauthoryear{Sigmund}{Sigmund}{2009}]%
        {sigmund2009manufacturing}
\bibfield{author}{\bibinfo{person}{Ole Sigmund}.}
  \bibinfo{year}{2009}\natexlab{}.
\newblock \showarticletitle{Manufacturing tolerant topology optimization}.
\newblock \bibinfo{journal}{\emph{Acta Mechanica Sinica}} \bibinfo{volume}{25},
  \bibinfo{number}{2} (\bibinfo{year}{2009}), \bibinfo{pages}{227--239}.
\newblock


\bibitem[\protect\citeauthoryear{Sumin, Rittig, Babaei, Nindel, Wilkie, Didyk,
  Bickel, K\u{r}iv\'{a}nek, Myszkowski, and Weyrich}{Sumin
  et~al\mbox{.}}{2019}]%
        {sumin2019geometry}
\bibfield{author}{\bibinfo{person}{Denis Sumin}, \bibinfo{person}{Tobias
  Rittig}, \bibinfo{person}{Vahid Babaei}, \bibinfo{person}{Thomas Nindel},
  \bibinfo{person}{Alexander Wilkie}, \bibinfo{person}{Piotr Didyk},
  \bibinfo{person}{Bernd Bickel}, \bibinfo{person}{Jaroslav K\u{r}iv\'{a}nek},
  \bibinfo{person}{Karol Myszkowski}, {and} \bibinfo{person}{Tim Weyrich}.}
  \bibinfo{year}{2019}\natexlab{}.
\newblock \showarticletitle{Geometry-aware scattering compensation for 3D
  printing}.
\newblock \bibinfo{journal}{\emph{ACM Trans. Graph.}} \bibinfo{volume}{38},
  \bibinfo{number}{4} (\bibinfo{year}{2019}).
\newblock


\bibitem[\protect\citeauthoryear{Sun, Xue, Rusinkiewicz, and Adams}{Sun
  et~al\mbox{.}}{2021}]%
        {sun2021amortized}
\bibfield{author}{\bibinfo{person}{Xingyuan Sun}, \bibinfo{person}{Tianju Xue},
  \bibinfo{person}{Szymon~M Rusinkiewicz}, {and} \bibinfo{person}{Ryan~P
  Adams}.} \bibinfo{year}{2021}\natexlab{}.
\newblock \showarticletitle{Amortized Synthesis of Constrained Configurations
  Using a Differentiable Surrogate}.
\newblock \bibinfo{journal}{\emph{arXiv preprint arXiv:2106.09019}}
  (\bibinfo{year}{2021}).
\newblock


\bibitem[\protect\citeauthoryear{Tjeng, Xiao, and Tedrake}{Tjeng
  et~al\mbox{.}}{2019}]%
        {tjeng2018evaluating}
\bibfield{author}{\bibinfo{person}{Vincent Tjeng}, \bibinfo{person}{Kai~Y.
  Xiao}, {and} \bibinfo{person}{Russ Tedrake}.}
  \bibinfo{year}{2019}\natexlab{}.
\newblock \showarticletitle{Evaluating Robustness of Neural Networks with Mixed
  Integer Programming}. In \bibinfo{booktitle}{\emph{International Conference
  on Learning Representations}}.
\newblock
\urldef\tempurl%
\url{https://openreview.net/forum?id=HyGIdiRqtm}
\showURL{%
\tempurl}


\bibitem[\protect\citeauthoryear{Tymms, Wang, and Zorin}{Tymms
  et~al\mbox{.}}{2020}]%
        {tymms2020appearance}
\bibfield{author}{\bibinfo{person}{Chelsea Tymms}, \bibinfo{person}{Siqi Wang},
  {and} \bibinfo{person}{Denis Zorin}.} \bibinfo{year}{2020}\natexlab{}.
\newblock \showarticletitle{Appearance-preserving tactile optimization}.
\newblock \bibinfo{journal}{\emph{ACM Transactions on Graphics (TOG)}}
  \bibinfo{volume}{39}, \bibinfo{number}{6} (\bibinfo{year}{2020}),
  \bibinfo{pages}{1--16}.
\newblock


\bibitem[\protect\citeauthoryear{Vielma}{Vielma}{2015}]%
        {vielma2015mixed}
\bibfield{author}{\bibinfo{person}{Juan~Pablo Vielma}.}
  \bibinfo{year}{2015}\natexlab{}.
\newblock \showarticletitle{Mixed integer linear programming formulation
  techniques}.
\newblock \bibinfo{journal}{\emph{Siam Review}} \bibinfo{volume}{57},
  \bibinfo{number}{1} (\bibinfo{year}{2015}), \bibinfo{pages}{3--57}.
\newblock


\bibitem[\protect\citeauthoryear{Xue, Beatson, Adriaenssens, and Adams}{Xue
  et~al\mbox{.}}{2020}]%
        {xue2020amortized}
\bibfield{author}{\bibinfo{person}{Tianju Xue}, \bibinfo{person}{Alex Beatson},
  \bibinfo{person}{Sigrid Adriaenssens}, {and} \bibinfo{person}{Ryan Adams}.}
  \bibinfo{year}{2020}\natexlab{}.
\newblock \showarticletitle{Amortized finite element analysis for fast
  PDE-constrained optimization}. In \bibinfo{booktitle}{\emph{International
  Conference on Machine Learning}}. PMLR, \bibinfo{pages}{10638--10647}.
\newblock


\bibitem[\protect\citeauthoryear{Yule}{Yule}{1967}]%
        {yule1967principles}
\bibfield{author}{\bibinfo{person}{John~AC Yule}.}
  \bibinfo{year}{1967}\natexlab{}.
\newblock \bibinfo{booktitle}{\emph{Principles of color reproduction: applied
  to photomechanical reproduction, color photography, and the ink, paper, and
  other related industries}}.
\newblock \bibinfo{publisher}{Wiley New York}.
\newblock


\bibitem[\protect\citeauthoryear{Zhu, Dave, Lipson, and Zheng}{Zhu
  et~al\mbox{.}}{2020}]%
        {zhu2020inverse}
\bibfield{author}{\bibinfo{person}{Ziwei Zhu}, \bibinfo{person}{Utsav~D Dave},
  \bibinfo{person}{Michal Lipson}, {and} \bibinfo{person}{Changxi Zheng}.}
  \bibinfo{year}{2020}\natexlab{}.
\newblock \showarticletitle{Inverse geometric design of fabrication-robust
  nanophotonic waveguides}. In \bibinfo{booktitle}{\emph{2020 Conference on
  Lasers and Electro-Optics (CLEO)}}. IEEE, \bibinfo{pages}{1--2}.
\newblock


\end{thebibliography}

\end{document}